\documentclass[conference]{IEEEtran}
\usepackage{balance}

\usepackage{enumitem}

\usepackage{cite}
\usepackage{xcolor}

\usepackage[font=small,skip=0pt]{caption}
\usepackage{graphicx}
\usepackage{subfigure}
\usepackage{amsfonts}

\usepackage{algorithm}

\makeatletter                                                                                                      
\newif\if@restonecol
\makeatother

\usepackage[linesnumbered,ruled,noresetcount,algo2e]{algorithm2e}
\SetAlFnt{\small} 

\usepackage{algorithmicx}
\usepackage[noend]{algpseudocode}
\usepackage{algpascal}

\usepackage{wrapfig}
\usepackage{lipsum} 

\usepackage{mathtools}

\usepackage{color}
\definecolor{boxshade}{gray}{0.2}

\newcommand{\cut}[1]{{}}

\newtheorem{definition}{\textbf{Definition}}

\def\BibTeX{{\rm B\kern-.05em{\sc i\kern-.025em b}\kern-.08em
    T\kern-.1667em\lower.7ex\hbox{E}\kern-.125emX}}

\begin{document}


%
\title{WedgeChain: A Trusted Edge-Cloud Store\\With Asynchronous (Lazy) Trust}


\author{
Faisal Nawab\\
University of California, Santa Cruz\\
Santa Cruz, CA 95060\\
fnawab@ucsc.edu
}

\maketitle


\begin{abstract}
\begin{sloppypar}
\cut{
Emerging applications in edge computing, Internet of Things (IoT),
and cloud computing, produce data that needs to be processed closer
to data sources at the edge to support real-time processing. and/or avoid
the need of sending huge amounts of data to the cloud. To enable
these applications, we
}
We propose WedgeChain, a data store that spans
both edge and cloud nodes (an edge-cloud system). WedgeChain consists
of a logging layer and a data indexing layer.
%
%
In this study, we encounter two challenges: (1)~edge nodes are
untrusted and potentially malicious, and (2)~edge-cloud coordination
is expensive.  WedgeChain tackles these challenges by the following
proposals: \textbf{(1)~Lazy (asynchronous) certification:} where data
is committed at the untrusted edge and then lazily certified at the
cloud node. 
This lazy certification method takes advantage of the observation
that an untrusted edge node is unlikely to act maliciously if it
knows it will be detected (and punished) eventually. Our lazy
certification method guarantees that malicious acts (\emph{i.e.},
lying) are eventually detected.
\textbf{(2)~Data-free certification:} our lazy certification method
only needs to send digests of data to the cloud---instead of sending
\emph{all} data to the cloud---which enables saving network and cloud
resources and reduce costs.
\textbf{(3)~LSMerkle:} we extend a trusted index
(mLSM~\cite{raju2018mlsm}) to enable indexing data at the edge while
utilizing lazy and data-free certification.

\end{sloppypar}

\end{abstract}

\section{Introduction}\label{sec:intro}

\cut{
Emerging applications in the areas of edge computing and Internet of
Things~(IoT) are promising to collect, store, and analyze huge
amounts of data.
}
To support processing this huge volume of data in
edge and IoT applications, the data management solution must be
capable of fast data ingestion at the edge---closer to data sources.
This is critical for two reasons: (1) many applications require
real-time processing---such as interactive mobile applications and
time-critical processing in Industry 4.0 and autonomous vehicles. (2)
many applications---such as large-scale video analytics and smart
city applications---produce huge amounts of data at a large number of
locations that would lead to increased costs for cloud communication
and computation.
\cut{
, \emph{i.e.}, public cloud providers charge a cost
for data transfer between the cloud and the Internet in addition to
the costs involved in storing and processing the data on virtual
machines. Doing more storage and processing at the edge would reduce
the costs involved in using cloud resources. 
}

However, processing data at the edge is complicated by the following
challenges: (1)~Edge nodes are untrusted. This is because the edge
node might be operated by a third-party provider {\color{black}outside
of the administrative domain of the} data
owner. This can also be due to
inexpensive or unmanaged edge nodes being more susceptible to
malicious breaches.
(2)~Edge-cloud coordination is expensive in terms of latency
(round-trip times are in the order of 100s of milliseconds to
seconds) and bandwidth (applications pay for data transfer costs
between the data center and the Internet).  This means that relying
on trusted nodes---in {\color{black}a trusted cloud or private cloud
in the administrative domain of the owner}---to authenticate the data is
expensive, and thus must be left out of the execution path of
requests and only utilized for asynchronous tasks. 

\cut{
Blockchain solutions offer an opportunity to provide a trusted,
decentralized data infrastructure for wide-area edge applications.
However, permissionless variants are expensive and incur a large
latency~\cite{nakamoto2008bitcoin}. On the other hand, permissioned
variants of
blockchain (that assumes that the participants are known and
authenticated~\cite{androulaki2018hyperledger,quorum,chain,parity,ripple,DBLP:conf/ndss/Al-BassamSBHD18})
can be divided into two categories: (1)~Byzantine-based variants:
those types depend on a byzantine fault-tolerance (BFT)
protocol~\cite{castro1999practical,kotla2007zyzzyva,cowling2006hq,
abd2005fault,kotla2004high,yin2003separating,li2007beyond,amir2010steward}
to add entries to the blockchain. BFT protocols, however, are
prohibitively expensive and face many practical challenges in real
deployments.  (2)~Variants based on benign (non-byzantine)
consensus~\cite{L98,L01,lamport2006fast}: these variants perform
well.  However, they do not safeguard from malicious activity of
untrusted edge nodes.
}


\begin{sloppypar}
In this paper, we propose \emph{WedgeChain}, an edge-cloud data store
that provides both logging and indexing of data for edge and IoT
applications. WedgeChain enables both an efficient and trusted data
storage and access.
{\color{black}In WedgeChain, the system model consists of authenticated clients
that produce data and send signed copies of the data to untrusted
edge nodes. The edge nodes service data access requests in
coordination with a trusted {private} cloud node that ensures that untrusted
edge nodes are not acting maliciously by providing an inconsistent
view of data\footnote{{\color{black}In the remainder of the paper, the
terms cloud and trusted cloud refer to a private cloud that is in the
administrative domain of the application owner.}}}
In WedgeChain, we propose the following three
features:

\textbf{(1)~Lazy (asynchronous) certification} enables committing
directly at the untrusted edge node and then asynchronously verifying
{\color{black}with the cloud node} that an edge node did not act
maliciously.
{\color{black}Specifically, the role of the cloud node is to prevent edge nodes from giving inconsistent views of the system to different clients.}
If an edge node is caught {\color{black} giving inconsistent views}, then it is punished. The way
WedgeChain implements this feature is by making the untrusted edge
node provide a signed message to the client that the data is
committed. This signed message is used by the client to prove that
the edge node lied in case the data was not actually committed. Our
observation is that in edge-cloud environments, nodes identities
are known (\emph{e.g.}, an edge node belongs to an IT department).
Therefore, an untrusted edge node would not act
maliciously if it knows that it will be caught and punished. The
punishment should be severe enough to outweigh the benefit of the
malicious act.

\textbf{(2)~Data-free certification} allows the certification at the
cloud to be performed using the data's digest which is smaller than
the data being certified. This allows
reducing the size of edge-cloud communication. This is possible
because agreement on the digest of data translates to agreement on
the data itself if the digest is a one-way hash function.

\textbf{(3)~LSMerkle} implements a fast-ingestion trusted index at edge
nodes that utilizes the lazy and data-free certification strategies of
WedgeChain.
\cut{
Existing trusted data indexing structures, such as
Merkle Trees~\cite{DBLP:conf/sp/Merkle80} cannot be integrated
efficiently to the WedgeChain model. This is because any change in
the index leads to a change to large parts of the Merkle Tree.
}
%
We integrate a new kind of indexing
structures---called mLSM~\cite{raju2018mlsm}---that combines the
design features of LSM trees~\cite{o1996log,luo2018lsm} (used for
high-velocity ingestion) and Merkle
trees~\cite{DBLP:conf/sp/Merkle80} (used for trusted data storage and
access.)
{\color{black}
LSMerkle uses mLSM as a data structure, replacing the memory component with a WedgeChain log/buffer and building the edge-cloud protocol around it to update and compact the mLSM structure in cooperation with the cloud node.
}
{\color{black}(This integration is
needed for key-value operations only; WedgeChain logging data
operations do not require the integration of a Merkle tree structure.)}
%

%

\cut{
\begin{sloppypar}
To summarize, WedgeChain consists of two components, a logging
component (Section~\ref{sec:wedgechain}) and a data index called
LSMerkle (Section~\ref{sec:lsmerkle}). The novel contributions in
WedgeChain's design are the following: (1)~lazy (Asynchronous) trust
verification, and (2)~integrating a trusted fast-ingestion data
indexing into lazy trust verification. 
}
\cut{
\begin{itemize}
  \item[1.] \emph{Asynchronous Trust Verification: } WedgeChain
allows the client and untrusted edge to update the state of the data
immediately while the verification process completes asynchronously
at the cloud node. Additionally, adding blocks to the edge node can
be verified by the cloud node without sending the whole block.
Rather, only the digest of the block is verified. WedgeChain design
ensures that verifying the digest is sufficient to verify the data on
the edge without having to actually send it to the cloud node.
  \item[2.] \emph{A fast-ingestion trusted index: }
LSMerkle integrates fast-ingestion index structures---such as
mLSM~\cite{raju2018mlsm}---into asynchronous trust verification.
This integration allows LSMerkle
to ingest data fast while enabling trusted data access. 
\end{itemize}
}
In the rest of this paper, we present background in
Section~\ref{sec:background}. Then, we present the model and design of
WedgeChain and LSMerkle in Sections~\ref{sec:model}
to~\ref{sec:lsmerkle}.  An experimental evaluation is presented in
Section~\ref{sec:eval} followed by a related work discussion in
Section~\ref{sec:related}. The paper concludes in
Section~\ref{sec:conclusion}.
\end{sloppypar}

\section{Background}
\label{sec:background}

\cut{
In this section, we introduce the target use cases and motivation of
edge-cloud data management and WedgeChain. Then, we introduce
relevant technologies that we use in WedgeChain. Finally, we present
a baseline solution for logging and data indexing in our target use
case and outline its drawbacks and limitations.  In the rest of the
paper, we present WedgeChain which overcomes the limitations of the
baseline solution.
}

\subsection{Target Use Cases}
\label{sub:target}

We target edge, IoT, and mobile applications where data is generated
in huge volumes and/or have workloads with real-time requirements.
\cut{
To support huge volume and real-time workloads, data must be stored
and processed at the edge. Transferring, storing and processing data
at the cloud incur a significant cost. For example, Amazon AWS charges
at least \$0.05 per GB of transferred data. If we consider a video
analytics application where only 1000 users are sharing their
720p-quality video streams, the cost of only transferring this amount
of data ranges between 10s to 100s of thousands of dollars per month
depending on the encoding and compression techniques.
}
%
Applications with real-time requirements, sending data to be
processed to {\color{black}a potentially faraway cloud node} is
infeasible as it can take hundreds of milliseconds to seconds to
receive a response (not counting the time to process the data). For
example, interactive mobile applications---such as Virtual and
Augmented Reality-based applications---require a latency of only tens
of milliseconds, which gives enough time to process the frames and
leave no time for wide-area communication.  This is also the case for
mission-critical applications in Industry 4.0 and autonomous
vehicles.

In this paper, we use the term \emph{edge} to represent any type of
resources that are close to users and sources of data. Edge devices
range from the client's personal devices (\emph{e.g.}, a router or
cluster of nodes in a building or university campus) to third-party
providers of edge data center technology, such as micro and mobile
data centers. The range of these resources are sometimes referred to
as mobile-edge computing, fog computing, and edge computing.

The challenge with leveraging edge infrastructure is that edge
resources---in many cases---are not 
{\color{black}in the same control domain as the application owner and
client} and thus might be untrusted.
{\color{black}
For example, consider a smart traffic application where a state
government is monitoring traffic to provide better routes and traffic
signals to vehicles and traffic controls such as traffic lights and
ramp meters. The data in this application includes information from
sensors and cameras that are placed around the city as well as
government-owned or contracted public transport.  The state
government (the owner of the application) has access to the
government data center that is faraway from the city (it is typical
for data centers to be placed in remote areas.) Therefore, the
application utilize edge resources at the city to enable fast
response to traffic conditions (\emph{e.g.}, reacting swiftly to a
traffic accident to reroute and change the flow in ramps).  

Although the application owner might have access to data sources
and traffic sensors around the city, it does not have compute
resources that are capable of data storage and processing at the
scale of this application. Therefore, it utilizes edge machines that
are operated by other entities outside of the owner's administrative
domain. This includes one or more of the following: (1)~third-party
edge service providers that rent out compute resources close to the
city. (2)~Independent contractors such as private transportation
companies that may integrate their vehicles with edge resources to
act as mobile edge resources. Both these types of edge resources
(edge service providers and independent contractors) are untrusted by
the application owner. Therefore, there is a need to maintain the
integrity of the data.  

}

\cut{
For example, consider a smart city application where sensors around a
city are generating data and the city leverages a public cloud
provider such as Amazon AWS for the data infrastructure and cloud
solutions. To enable edge-cloud computing, the city or the public
cloud provider would contract with edge resource providers to store
and process data at the edge. These edge resource providers can range
from telecommunication companies to private third-party contractors.
}

This direction of utilizing edge resource providers is now starting
to manifest {\color{black}
as services provided by various entities including public
cloud providers.} For
example, Amazon AWS services such as Amazon
Wavelength~\cite{wavelength} are partnering with telecommunication
providers such as Verizon, Vodafone, and SK telecom to allow Amazon
AWS cloud compute to be hosted on their edge resources (e.g.,
cellular towers and 5G infrastructure).  Similarly, Microsoft Azure
partnered with AT\&T for the same purpose~\cite{msatt}.
{\color{black}(This corresponds to edge providers in the example above.)}
\cut{
As these
partnerships grow to include more third-party organizations, the
problem of untrusted edge devices exacerbates.
}
Similarly, other cloud services (such as Amazon AWS IoT
Greengrass~\cite{greengrass}) allow customers to deploy cloud
functions on the customer's edge devices.  In these types of
applications, operations on user's edge devices must be performed in
a trusted manner. {\color{black}(This corresponds to independent
contractors in the example above.)}

Figure~\ref{fig:usecase}(a) shows an example of an edge-cloud
deployment. Clients access the data or generate data from their
devices and send data to the WedgeChain data system that
consists of edge nodes and a cloud node.  There are two types of data
requests. The first is for data logging and streaming requests, which
consists of \textsf{add()} and \textsf{read()} operations.  The other
type of data access is for key-value requests, which consists of
\textsf{put()} and \textsf{get()} key-value operations.

Each \emph{edge node} handles the storage and processing for a subset
of the clients (\emph{i.e.}, a partition of the data).
{\color{black}Specifically, each client is associated with a single
partition/edge node. Thus, finding the data that pertains to a client
is done by directing the request to its corresponding client.
Due to the spatial locality of edge applications, we focus on single-partition operations in this work.} Edge nodes can
be edge and micro datacenters or user devices. The \emph{cloud node}
maintains the rest of the application's data (and potentially a
backup of a subset of the data on edge nodes). It also helps edge
nodes in certifying data and running maintenance tasks. 

\begin{figure}[!t]
\begin{center}
\label{fig:usecase}
\includegraphics[scale=.7]{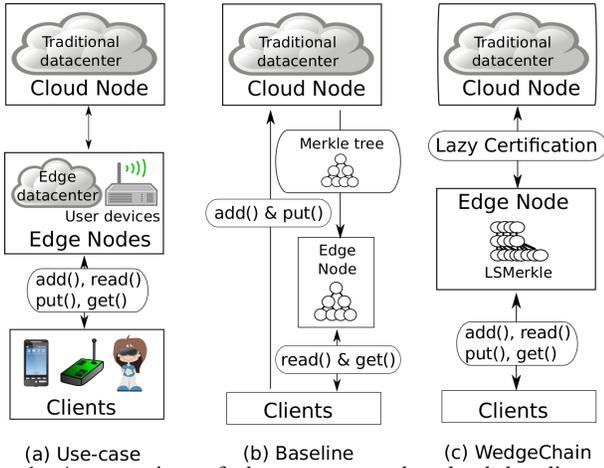}
\caption{An overview of the use-case, edge-cloud baseline, and architecture of
WedgeChain.}
\label{fig:usecase}
\end{center}
\end{figure}

\subsection{Relevant Technologies}
\label{sub:merkle}

\cut{
We build upon two key technologies, LSM and Merkle
trees.
}

\subsubsection{LSM Trees}
Log-Structured Merge (LSM) Trees are designed to support fast
ingestion of data.
\cut{
Many other indexing data structures, such as the
B-Tree, update data in-place, which requires traversing the tree for
each update. This is infeasible when it is desired to optimize for
fast ingestion.
}
LSM trees
batch updates in pages and merge them with the rest of the data
later. This moves merging the data out of the execution path of
updates, hence making ingestion more efficient. There are many LSM
tree variants~\cite{o1996log,luo2018lsm}. In general, the
tree is structured into $L$ levels. Level 0 is where new pages are
appended and is maintained in main memory. Once the number of pages
in Level 0 exceeds a threshold, then the pages are merged with the
next level, Level 1. Levels 1 and higher are persisted in storage.
Each level has a threshold, when exceeded, pages are merged with the
next level. The details of these operations and structure vary across
designs~\cite{luo2018lsm}.

\subsubsection{Merkle Trees}
\label{sub:merkle-bg}
Merkle trees~\cite{DBLP:conf/sp/Merkle80}
allow an untrusted node to serve data in a trusted way.
Specifically, it allows the untrusted node to provide a proof of the
authenticity of the data that {\color{black}are} originally generated and signed by
a trusted node. The way Merkle trees are usually designed
is by dividing the data into pages. Then, Each page is hashed with a
cryptographic hashing function.  The hashes of the pages
represent the leaves of the Merkle tree. Then, each pair of leaf
nodes are hashed to construct a node at the next level of the Merkle
tree. This is continued until there is only a single node in a level.
This node is called the \emph{Merkle root}. A trusted entity
(\emph{e.g.}, the trusted cloud node in our case) signs the
Merkle root to certify the authenticity of the data. The untrusted
node uses this signed Merkle root to provide a proof to clients that
the data is authentic.
\cut{
For example, if a client asks for a value.
The untrusted node sends back the page that
contains the value in addition to the siblings of all the Merkle tree
nodes in the route to the Merkle Root. Also, it sends the signed
Merkle root. When a client receives the response, it uses the
received page and the sibling Merkle tree nodes to construct the
Merkle root of the tree. Then, it compares the Merkle root it derived
with the Merkle root signed by the trusted node. If they match, then
the client can trust that
the data is authentic. The reason for this, is that it is
computationally infeasible for a malicious node to fabricate a page
that leads to constructing the same Merkle root that is signed by the
trusted node.
}
{\color{black}In this work, we leverage Authenticated Data Structures
such as Merkle Trees to enable key-value access to data from
untrusted edge nodes. However, these structures are not needed for
logging data operations which utilize lazy (asynchronous) trust
directly.}

\subsection{Baseline Solutions and Their Drawbacks}
\label{sub:baseline}

Given the use case and relevant technologies introduced in this
section, it is possible to come up with a straight-forward solution
for edge-cloud data management with untrusted edge nodes. We call
this \emph{edge-baseline} and show it in Figure~\ref{fig:usecase}(b).
Specifically, clients send their \text{add()} and \textsf{put()}
requests to the (trusted) cloud node.  Then, the cloud node
regenerates the Merkle tree to account for the new updates, sign the
Merkle root, and send the Merkle tree to the untrusted edge node.
This enables the edge node to serve data access requests by using the
signed Merkle root as proof of the data's authenticity.

However, this straight-forward solution has a drawback. Whenever data
needs to be logged (using \textsf{add}) or inserted into the data
structure (using \textsf{put}), the cloud node is in the path of
execution. 
%
\cut{
In the case of \textsf{add} calls, the block must be sent to the
cloud node to be signed. 
%
In the case of \textsf{put} calls, the Merkle tree must be
recomputed and signed by the cloud node.
%

Another drawback is related to how the Merkle tree needs to be
reconstructed for every update (or group of updates). One
update to the Merkle tree would result in the need to recompute the
hashes along the path from one leaf node to the Merkle root. In the
typical case, however, updates to the index are batched and may span
many pages in the Merkle tree. This makes it likely that every batch
of updates would lead to recomputing a large portion of the Merkle
tree.
}
Our proposal WedgeChain overcomes these limitation by
employing a lazy (asynchronous) certification strategy that takes the
cloud node out of the execution path of \textsf{add} and \textsf{put}
operations---see Figure~\ref{fig:usecase}(c). In WedgeChain, data
access requests are served immediately from the edge nodes without
having to wait for the cloud node to certify the data. To make sure
that the edge node does not lie, the edge node provides a temporary
proof in its response. This temporary proof can be used later by the
client to detect  if the edge node lied. If the
edge node lied to the client, then the client can use the temporary
proof it received to prove that the edge node is malicious and
thus {\color{black}is} able to punish the malicious node. 


\cut{
In the evaluation section, we compare with edge-baseline that we
introduced above. We also compare with a baseline that we call
\emph{Cloud-only}, which utilizes the cloud node without
involving any edge nodes.
}

{\color{black}
\subsection{Security Model Assumptions}
\label{sub:assumptions}

Our lazy certification method is enabled by observing that some
security model characteristics of existing systems can be relaxed in
applications of edge-cloud systems with a hybrid trust model.
Specifically, we make the following assumptions about the security
model (and how they are reflected in the smart traffic application we
presented above):\\
1. The application owner can enforce a punishment that would deter
untrusted edge nodes from committing malicious acts. In the smart
traffic application, for example, this assumption can hold by
enforcing a monetary and/or legal punishment. For both edge service
providers and independent contractors, since they are known entities,
the application owner can enforce the punishment. \\
2. The application can prevent an untrusted node that acted
maliciously before from reentering as an edge node. In the smart
traffic application, because the real identities of both edge
providers and independent contractors are knows, and they cannot
fabricate new identities, the application owner can prevent their
reentry. \\
3. Malicious acts cannot lead to catastrophic consequences. This
condition can be trivially satisfied by handling critical operation
that can be catastrophic at the cloud. The definition of
catastrophic depends on the application. In our smart traffic
application, for instance, destroying the data at an edge location
might not be deemed catastrophic as the application state depends on
the collective information of a large number of sensors/cameras and
the small potential of nodes acting maliciously (due to assumption 1
above) outweighs the inaccuracy and potential lost of information.

}

\section{{\color{black}System and Data Model}}
\label{sec:model}

The system consists of three types of nodes: (1)~cloud nodes that are
trusted (non-byzantine). Each cloud node can be backed by
a high-availability cluster for availability. For ease of exposition,
however, we assume that
there is one cloud node.
{\color{black}The role of the cloud node is to ensure that edge nodes
are not providing an inconsistent view of data to clients.}
 (2)~edge nodes that are not trusted. An edge
node receives data from clients and stores it locally in the form of
a log {\color{black}or index}. It also receives requests to access stored data from clients.
(3)~clients are {\color{black}authenticated} nodes that generate and consume data from edge nodes
and devices. {\color{black}The generated data is signed and sent to
edge nodes for processing.}

Each edge node maintains a log that pertains to
a subset of {\color{black}clients (edge devices)}. For example, in an
application with IoT sensors, each edge node maintains the data
generated by a set of the IoT sensors {\color{black}(\emph{i.e.}, clients)}.
{\color{black}Also, each client belong to a single partition on a
single edge node.}
Each block is a batch of data
entries. Clients may read a block by issuing a request with the
block's id to the edge node. 
{\color{black}Block ids are unique monotonic numbers that are assigned
by the edge node (the ids are unique relative to an edge nodes, but
are not unique across edge nodes.)}
In addition to the log, each node may maintain an index data
structure. We present more details about the index data structure in
Section~\ref{sec:lsmerkle}.

\section{{\color{black}WedgeChain Logging}}
\label{sec:wedgechain}

\begin{sloppypar}
In this section, we present WedgeChain
and the detailed design of the logging component.
{\color{black}In
Section~\ref{sec:lsmerkle}, we present the indexing component.}
\end{sloppypar}

\subsection{{\color{black}Logging Interface}}
\label{sub:wedgechain-model}
The edge node's interface consists of the following calls (all
message exchanges are signed by the sender):
\begin{itemize}
  \item \textsf{add(in: entry, out: bid, (optional) block)}: this
call adds an entry to the next block at the edge. The edge node
returns the block id ($bid$) that contains the entry. If requested,
the edge node returns the newly formed block that contains the entry.
  \item \textsf{read(in: bid, out: block, bid, proof)}: this function
takes a block id number as input and returns the corresponding block
in addition to a proof of the authenticity of the block. This proof
might be either \emph{(1)~in-progress (Phase I)} or
\emph{(2)~final (Phase II)}. More
details about proofs and commit phases later in the section. 
\end{itemize}
{\color{black}Each of these logging operations is performed on a single
block, independent of prior blocks. The cloud node ensures that
untrusted edge nodes are not giving an inconsistent view of the
blocks. Because logging operations operate at the level of single
blocks, the detection of malicious behavior by the cloud can operate
at the level of single independent blocks as well. This limits the
type of stateful operations running on the log. For this reason, we
also present a key-value operations that maintain state across blocks
in Section~\ref{sec:lsmerkle}.}

\subsection{WedgeChain Overview}

\textbf{Guarantees.}
The main goal of WedgeChain is to support adding to and reading from
the edge node's log while guaranteeing \emph{validity} and
\emph{agreement}.  Validity
is a guarantee that an entry in the log is one that has been
proposed by a client.  Agreement is a guarantee that any two nodes
reading the same block will observe the same content.

\textbf{Lazy (Asynchronous) Certification.}
Lazy certification distinguishes between two types of commitments:
\emph{initial commit (Phase I Commit)} and \emph{final commit (Phase
II Commit)}. Initial commit is the commitment done without involving
the trusted cloud node. Instead, the untrusted edge node provides a
temporary proof to the client. This temporary proof can be used by
the client later to prove that the edge node promised to add the
entry to a specific block. Therefore, a malicious edge node can be
detected and punished. The final commit phase is when the trusted
cloud node authenticates the request either ensuring that the edge
node did not lie in its response or proving that it lied and should
be punished.

\textbf{Initial (Phase I) Commit is Sufficient to Make Progress.}
The ability to detect malicious behavior allows a client to commit
immediately and make progress after Phase I commit. This is because
an untrusted edge node does not have an incentive to lie since it
knows that it will eventually be detected. This assumes that the
harm of the penalties/punishments that would be applied to a
malicious edge node outweigh the benefit of the malicious activity.

\textbf{Coordination Pattern (Phase I Commit).}
In the rest of this section, we cover how WedgeChain enables adding
and reading from a single edge node's log. Consider a scenario
with a client $c$, an edge node $e$, and a cloud node $\mathcal{L}$.
The client $c$ can be an IoT sensor or edge device that generates
data continuously.  Assume that $c$ sends all its data to $e$ for it
to be stored in its log. Client $c$ sends an \textsf{add} request to
$e$ to add entries.  Upon receiving the \textsf{add} request, $e$
batches the client's sent entry, $m$, with other requests to be
committed as part of the next block. Once a block is ready, the block
id and the block containing the entry $m$ are returned to the
client $c$. At this time, the entry and block are Phase I Committed. 

\textbf{Coordination Pattern (Phase II Commit).}
Concurrently, the edge node $e$ sends the digest of the block (that
contains both the content and the block id) to the cloud node
$\mathcal{L}$. The digest must be constructed using a one-way hash
function.  Then, $\mathcal{L}$ sends back a message that
contains the signed digest of block $bid$ if it is the first time it
receives a digest for block $bid$. Otherwise, the cloud node detects
a malicious activity and rejects the request. The signed message from
$\mathcal{L}$ acts as a certification that this is the block digest
that is committed. The cloud node also maintains the digests of all
committed blocks of edge nodes. At this time, the entry and the block
are Phase II Committed.

\textbf{Data-Free Coordination.}
Note that the edge node only needs to send the digest (constructed
with a one-way hash function) to the cloud
node during Phase II Commit. This is beneficial because it reduces
the edge-cloud communication overhead. Data-free coordination is
possible because the digest is used as a proxy of the actual content
of the data. Therefore, if all clients agree on the committed digest $d$
of a block $B$, then they also agree on the content of the block $B$.

\textbf{Certification.}
A digest that is accepted and signed by the cloud node is called a
\emph{certified digest} and the corresponding block is called a
\emph{certified block}. A client can ensure that its entry is Phase
II Committed by checking whether the block it received got certified
by the cloud node. This can be done by contacting the cloud node
directly or asking the edge node to forward the signed digest. This
certification also guarantees agreement, since an edge node cannot
certify two different blocks as Phase II committed with the same
block id.

\textbf{Reads.}
The signed digest is also used to certify reads. When a read request
is sent to the edge node, the edge node responds with the block and
the signed digest (denoted \textsf{proof} in the interface). The
client can then verify the authenticity of the block by computing the
digest and comparing it to the proof. For blocks that are not yet
certified by the cloud node, the edge node may utilize lazy
certification and send the block and an empty proof. The client will
get the certification from the cloud node eventually and can detect
whether the edge node was malicious, similar to the case of the
\textsf{add} interface. If the edge node lied in its response, then
the client can show the response to the read request as a proof and
thus punish the edge node. This ability to prove maliciousness would
deter malicious activity and enable clients to consider the
\textsf{add} committed with high certainty even before Phase II
Commitment.


\begin{figure}[t]
\centering
  \includegraphics[width=3.2in]{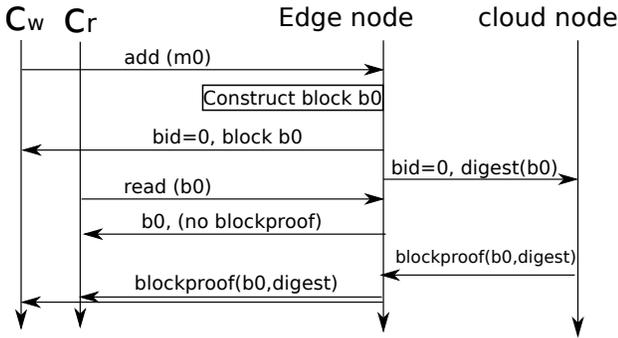}
  \caption{{\color{black}An example of the coordination necessary to add and read
blocks in WedgeChain.}}
  \label{fig:example}
\end{figure}

\subsection{Example}

{\color{black}
Consider a scenario (Figure~\ref{fig:example}) with two clients,
$c_w$ and $c_r$, an edge node, and a cloud node. 
Initially, $c_w$ sends the data entry $m_0$ to the edge node.  The
edge node creates a block $b_0$ with $m_0$ in its payload. Then, it
sends the signed block and its id back to $c_w$. Client $c_w$ uses
this response to terminates its Phase I Commit and continues
operation while lazy certification is performed in the background. 
Asynchronously, the edge node sends the digest of $b_0$ to the cloud
node to be certified (Note that only the digest need to be sent, not
the whole block.) While the edge node is waiting for the cloud node,
the other client, $c_r$, sends a request to read $b_0$. The edge node
responds with the content of $b_0$ but with no certification from the
cloud (called \emph{blockproof} in the figure). Client $c_r$ uses
this response to terminates its Phase I Commit and continues
operation. 
Afterward, the certification is sent from the cloud node to the edge
node for $b_0$. The edge node forwards the certification (called
blockproof in the figure) to both $c_w$ and $c_r$, which terminates
their Phase II Commit.

}

\cut{
To illustrate the operation of WedgeChain, consider a scenario with
two clients, $c_w$ and $c_r$, an edge node, and a cloud node. Client
$c_w$ writes its data to the edge node and $c_r$ reads data from the
edge node. Figure~\ref{fig:example} shows the communications in this
example. 

Initially, $c_w$ sends the data entry $m_0$ to the edge node.
The edge node creates a buffer for its first block, $b_0$, and adds
the entry to the buffer. Then, $c_w$ sends 9 more entries to the edge
nodes, $m_1$ to $m_9$. The edge node adds these entries to the
buffer. Assuming that the buffer threshold is met, the edge node
writes the contents of the buffer to block $b_0$. Then, it sends the
block and its id back to $c_w$. At this point, the entry $m_0$ is
Phase I committed.

Then, the edge node sends the digest to the cloud node to be
certified (note that only the digest was sent, not the whole block.)
While the edge node is waiting for the cloud node, the client $c_w$
sends more entries to the edge node that are written to block $b_1$
similar to how block $b_0$ was formed.  Afterwards, the certification
is now received from the cloud node and the edge node forwards it to
$c_w$. At this point, the entry $m_0$ is Phase II committed.

Now, the edge node has two blocks, one of which is certified. Let's
examine how reads are performed. Client $c_r$ begins by requesting to
read block $b_0$. The edge node responds with the block and the
signed digest that it got from the cloud node. Upon reception, $c_r$
computes the digest from the received block and compares it with the
signed digest. If they match, then this terminates the read process
and the read is considered Phase II committed.
Otherwise, a dispute process is invoked and sent to the cloud node. 

Next, $c_r$ sends a request to read block $b_1$. The edge node
responds with the block and an empty proof, since it did not receive
the signed certification from the cloud node yet. The client receives
the block, considers the read Phase I committed, and engages in the
lazy certification process. In this case,
$c_r$ computes a digest of the block and sends it to the cloud node.
The cloud node waits until it gets the certification request from the
edge node. If the digests match, then the cloud node responds with a
positive acknowledgment that represents the certification and the
read is then considered Phase II committed. Otherwise,
the cloud detects a violation and starts a dispute process to
punish the edge node.
}

\subsection{Algorithms}


\subsubsection{\textbf{Adding to the log}}

The following are the algorithms to add a block to the edge node's
log.

\alglanguage{pseudocode}
\begin{algorithm2e}[t]
\caption{Client algorithm to add an entry. }
\label{alg:add}
\begin{algorithmic}[1]
\State on AddNewEntry (in: entry) \{
\State \,\,\,\,\,Send \textsf{add(\$entry) to edge node}
\State \,\,\,\,\,$\$block, \$bid$ $\leftarrow$ edge
node response
\State \,\,\,\,\,Verify that the entry is in $\$block$
\State \,\,\,\,\,Mark $\$entry$ as Phase I Committed
\State \,\,\,\,\,Wait until \textsf{block-proof} is received 
\State \,\,\,\,\,$\$blockProof \leftarrow$ \textsf{block-proof}
message from cloud node 
\State \,\,\,\,\,Verify that $digest(\$block \& \$id) = \$blockProof$
\State \,\,\,\,\,Mark \textsf{add} as Phase II Committed
\State \}
\end{algorithmic}
\end{algorithm2e}

\textbf{Client algorithm (Algorithm~\ref{alg:add}).}
The client constructs an \textsf{add} message that contains the data
it wants to add to the log. In our model, the client---which
represents an IoT sensor or edge device---is authenticated. To trust
the \textsf{add} message, the client includes a signature. The client
sends the signed \textsf{add} message to the (untrusted) edge node.
Then, it waits until it hears a response from the edge node that
contains the contents and block id of the block that contains the
added entry. This response is signed from the edge node (This is
important since the client can use this signed response in the event
of a dispute to punish the malicious node.) The client verifies that
its entry---that corresponds to the \textsf{add} request---is part of
the block. 

After hearing the \textsf{add-response} message from the edge node
and verifying its contents, the client marks the corresponding
\textsf{add} request as a \emph{Phase I Commit}. This Phase I Commit
represents the following guarantee:
\begin{definition}
\textbf{(Phase I Commit Guarantee)}
If an entry is Phase I Committed in block $bid$, then that implies
that either (1)~the entry is part of block $bid$ or, otherwise,
(2)~the client can successfully prove that the edge node is malicious
and thus the edge node would be punished.
\end{definition}

Eventually, the client receives a \textsf{block-proof} message
from the cloud node---that might be forwarded by the edge node. The
\textsf{block-proof} message is signed by the cloud node to ensure
its authenticity. It contains the block id, $bid$, and its corresponding
digest. Upon receiving this message, the client marks the
\textsf{add} request as a \emph{Phase II Commit} which guarantees:
\begin{definition}
\textbf{(Phase II Commit Guarantee)}
If an entry is Phase II Committed in block $bid$, then this means
that the edge node cannot report another block for this block id as
Phase II Committed. Therefore, it is impossible for two clients
to disagree about the content of a block if their operations on it
were both Phase II Committed. 
\end{definition}

\textbf{Edge node algorithm.}
When an edge node receives an \textsf{add} request, it verifies the
authenticity of the message by checking the signature and that it
belongs to a certified client. Then, it adds it to a buffer. Once the
buffer is full, a new block is constructed with the entries in the
buffer and appended to the log. Then, the edge node constructs
an \textsf{add-response} message for each entry in the block (the
\textsf{add-response} messages can be aggregated and sent together if
they belong to the same client.) The \textsf{add-response} message is
signed by the edge node and includes the block and block id. These
messages are then sent to the corresponding clients.

After adding the block to the log, the edge node sends a
signed \textsf{block-certify} message to the cloud node that contains
the block id and block digest. The cloud node sends back a signed
\textsf{block-proof} message with the block id and digest. The edge
node forwards the \textsf{block-proof} message to all clients that
added entries in the corresponding block.

\textbf{Cloud node algorithm.}
The cloud node receives a signed \textsf{block-certify} request from
an edge node that contains the block id and digest. It verifies that
it did not hear any prior requests to certify a block with the same
block id. If it is the first, then it sends back the
\textsf{block-proof} message with the block id and digest. Otherwise,
it flags the edge node as malicious.

\subsubsection{\textbf{Reading from the log}}
The following are the algorithms to read a block.

\cut{
\alglanguage{pseudocode}
\begin{algorithm2e}[t]
\caption{Client algorithm to read a block. }
\label{alg:read}
\begin{algorithmic}[1]
\State on ReadBlock (in: bid) \{
\State \,\,\,\,\,Send \textsf{read(\$bid) to edge node}
\State \,\,\,\,\,$\$block, \$bid, \$digest$ $\leftarrow$ edge
node response
\State \,\,\,\,\,\textbf{if} $\$block$ = \textsf{null} \textbf{then}
\State \,\,\,\,\,\,\,\,\,\,return and exit // contact cloud node if necessary
\State \,\,\,\,\,Verify that \textsf{digest}($\$block$ \& $\$bid$) =
$\$digest$
\State \,\,\,\,\,Mark \textsf{read} as Phase I Committed
\State \,\,\,\,\,return $\$block$ to reader
\State \,\,\,\,\,Wait until \textsf{block-proof} is received 
\State \,\,\,\,\,$\$blockProof \leftarrow$ \textsf{block-proof}
message from cloud node 
\State \,\,\,\,\,Verify that $\$digest = \$blockProof$
\State \,\,\,\,\,Mark \textsf{read} as Phase II Committed
\State \}
\end{algorithmic}
\end{algorithm2e}
}

\textbf{Client algorithm.}
To read a block, a client sends a \textsf{read} request with the
block id that it wishes to read to the edge node. There are three
cases:
\begin{itemize}
\item[1.] \emph{The block is not available: }
The edge node responds with a signed message saying that
the block is not available. At this point, if the client is
suspicious that the edge node is malicious and lying about the
unavailability of the block, it can send a request to the cloud node
asking whether the block was reported.
\item[2.] \emph{Phase II Commit read:}  A signed response that
includes the block and a proof. The proof is a \textsf{block-proof}
message that has been signed by the cloud node. The client verifies
the \textsf{block-proof} and terminates the read.
\item[3.] \emph{Phase I Commit read: } A signed
response that includes the block but without a proof. In this case,
the client waits for a \textsf{block-proof} to be sent from the cloud
node. After receiving the response, but before receiving the
\textsf{block-proof}, the read is considered Phase I Committed. The
client can successfully dispute the read response in the case it
turns out that the edge node lied in its initial response. Once the
\textsf{block-proof} is received, the read is considered Phase II
Committed.
\end{itemize}

\textbf{Edge node algorithm.}
When an edge node receives a read request, it checks whether the
requested block is available. If it is not, then the edge node
responds negatively. Otherwise, the edge node responds with the
block. If a \textsf{block-proof} is available, then it is sent with
the block. Otherwise, an empty proof is sent. Eventually, when a
\textsf{block-proof} is received from the cloud node, the edge node
forwards it to the client.

\subsection{{\color{black}Security Threats}}
\label{sub:threats}

{\color{black}
\textbf{Replay attacks.}
A replay attack is performed by the malicious edge node repeating a
valid client request more than once. To overcome this attack,
existing techniques can be integrated without incurring extra
communication overhead to the cloud. The choice depends on the what
the application permits. Specifically, in many edge applications,
requests are idempotent which means that applying the request more
than once has the same effect as applying it once, \emph{e.g.}, a
sensor indicating that the temperature reading is $x$ at timestamp
$ts$ has the same effect when repeated. Generalization of this using
timestamps, session ids, and prior state (\emph{i.e.}, explicitly
defining the prior state in the request) can all be integrated from
the client-side without affecting WedgeChain. WedgeChain can also be
extended to provide support to make any arbitrary request idempotent.
This can be done by making each request signed by the client for a
specific log position. Specifically, the client first reserves a log
position via a round of messaging with the edge node. Then, the
client signs the request with the reserved log position. Because the
request is signed for a specific log position, any other client would
not accept the request if it is in another log position. This design
does not lead to extra edge-cloud communication. Also, the
reservations can be mandatory (the block waits for all reserved
requests) or best-effort (if some reserved requests are late, then
they are discarded, and the client has to do another reservation.)

\textbf{Omission attacks.}
A malicious edge node might respond negatively to a read request of a
log position that is actually filled (either to delay the response or
because data was maliciously destroyed). Minimizing the effect of
this omission attack can be performed by asynchronous gossip
propagation from the cloud node to clients (either through the edge
node or directly from the cloud node). These gossip messages are
signed by the cloud node with a timestamp and the log size as of that
timestamp. A client can use these gossip messages to know that all
log positions smaller than the log size are filled. This still leaves
the opportunity for omission attacks on recent data. The time-window
of this threat is a function of the frequency of gossip messages. (We
also discuss omission attacks as they pertain to key-value operations
in Section~\ref{sub:freshness}.)
}

\textbf{Disputes.}
A dispute can arise if the client discovers that the edge node has
lied in its response. There are malicious acts that can be detected
trivially, such as responding with a digest that does not match the
block or signing with the wrong signature. Other than these types
of malicious acts, an edge node might respond to an \textsf{add} or
\textsf{read} request with incorrect information that cannot be
immediately detected:
\begin{itemize}
  \item[1.] \textsf{add-response: } the edge node responds that the
entry is going to be in block $i$, but then the actual certified
block $i$ does not include the entry. 
  \item[2.] \textsf{read-response: } the edge node responds with
block $i$ and no proof, but it turns out the block is not the one
committed with id $i$.
\end{itemize}
In both cases, the client discovers the malicious act after the call
has entered Phase II Commit. Because the edge node lied about the
content of the block, it cannot provide the \textsf{block-proof}
message, since it must be signed from the cloud node. The client
waits for the \textsf{block-proof} message. If it does not
receive it for a predefined time threshold, it sends a request to the
cloud node with the block id and digest. The cloud node detects that
the digest does not match what is reported by the edge node. In such
a case, the edge node is punished. 

{\color{black}
A dispute can also be sent if an omission attack is detected (described above by using the gossip from the cloud node). A dispute message is sent to the cloud node to force the edge node to respond, and if it does not, then a punishment procedure starts.
}

{\color{black}
\textbf{Availability attacks} such as an edge node
that delays responding to messages (but responds correctly) to degrade the system performance
are more complicated and remain as an open problem.
}

\cut{
\subsection{Correctness Proofs}
?????????? XXXXXXXXXxxx
}



\section{LSMerkle Design}
\label{sec:lsmerkle}

In this section, we extend WedgeChain
with a data indexing structure that enables accessing the key-value
pairs in the log through a key-value interface of \textsf{get} and
\textsf{put} operations. 

Our proposal provides an index on top of WedgeChain that is both
efficient and trusted. This means that a potentially malicious edge
node can respond to client requests without having to involve the
cloud node. Our data index---called LSMerkle---builds upon
mLSM~\cite{raju2018mlsm} and extends it to work in the edge-cloud
environment of WedgeChain and to support lazy (asynchronous)
certification.
{\color{black} Specifically, LSMerkle uses mLSM as the data structure
and builds around it a protocol to coordinate with clients and the
cloud node the update and compaction operations as well as
integrating a WedgeChain log/buffer as the memory component to allow
Phase I Commit with asynchronous trust.}

\cut{
In the rest of this section, we present the system and data model, a
trivial baseline solution using Merkle Trees, and our more efficient
index structure, LSMerkle, that enables fast ingestion and trusted
data access on top of WedgeChain.
}

\subsection{System and Data Model}

The indexing structure extends WedgeChain. This means that it
inherits the system model of WedgeChain that consists of clients,
edge nodes and a cloud node. The LSMerkle tree is stored in the edge
node and its interface consists of:
\begin{itemize}[leftmargin=*]
 \item \textsf{put(IN: key, value, OUT: block, bid)}: {\color{black}
this request
takes a key-value pair and applies it to the index. The return values
are ones that correspond to the block where the key-value pair are
added.}
\cut{
This put request
acts as an \emph{upsert} operation---updating the key if it exists or
adding the key if it does not exist. The way this request is handled
is similar to the WedgeChain \textsf{add} interface. The effect of
this call is adding an entry that represents the operation in the
log. This is done in exactly the same way as \textsf{add}
calls. What is different is that there is also an additional update
to the index structure. Specifically, the index is updated with the
corresponding key and value pair.
}
 \item \textsf{get(IN: key, OUT: value, index\_proof)}:
This call returns the value of the requested key and a proof of the
authenticity of the response. We provide more
details about the index\_proof, but at a high-level it consists of
certified parts of the index that prove that the value is part of the
index. It also includes certifications similar to log \textsf{read}
Phase I and Phase II Commits.
\end{itemize}

\cut{
\subsection{Baseline Solution Using Existing Trusted Index Structures}
\label{sub:baseline2}

Using existing trusted data indexing structures (such as Merkle trees
and mLSM~\cite{raju2018mlsm}) lead to degrading the performance of
WedgeChain in edge-cloud environments. This is because current
trusted data indexing structures require a trusted node to construct
the index. Therefore, any update to the trusted index---in our
case---must go through the (trusted) cloud node. To show how this is
the case, in this section, we present a case study of using a Merkle
tree in WedgeChain and observe how all updates must synchronously go
through the cloud node. (Afterwards, we show how LSMerkle overcomes
this by integrating the index into WedgeChain and lazy certification.) 

\begin{sloppypar}
\textbf{Integrating a Merkle Tree into WedgeChain.}
The index data structure consists of the ordered set of $n$ keys
($k_0,\ldots,k_{n-1}$) divided into an ordered set of $m$ pages
($p_0,\ldots,p_{m-1}$).  These pages are used to construct a Merkle
Tree (see Section~\ref{sub:merkle} for details about Merkle trees).

When a new block is constructed, the client needs to send the block
to the trusted cloud node to update the Merkle tree.  The cloud node
updates the Merkle tree according to the \textsf{put} operations in
the block and sends the new Merkle tree to the edge node with the
signed Merkle root.
\end{sloppypar}

When a client reads a key from an edge node, the edge node sends the
value, the Merkle tree sibling path to the root to enable the client
to compute the corresponding Merkle root (see
Section~\ref{sub:merkle}), and the proof which is a signed Merkle
root from the cloud node. The client uses the signed Merkle root to
verify the authenticity of the response.

The problem in this straight-forward integration of Merkle trees into
WedgeChain is that the cloud node must be in the path of execution of
updating the Merkle tree. This makes the update latency proportional
to the round-trip time latency to the cloud node, which can be up to
100s of milliseconds to seconds. 

This limitation exists for any type of trusted index since updating
the trusted index must be done using a trusted node (\emph{e.g.,} the
cloud node in WedgeChain.) However, we observe that it is possible to
adapt trusted index structures to WedgeChain and enable them to
utilize lazy (asynchronous) certification. In the rest of this
section, we show how our index structure, LSMerkle, achieves this.
LSMerkle builds on mLSM~\cite{raju2018mlsm} as its design is amenable
to lazy certification. 
}



\subsection{LSMerkle Design}
\label{sub:lsmerkle-design}

We propose \emph{LSM Merkle tree} (LSMerkle), a data indexing
structure that can be integrated into WedgeChain and utilize lazy
(asynchronous) certification.  LSMerkle builds on
mLSM~\cite{raju2018mlsm} that combines techniques from LSM trees (for
fast ingestion) and Merkle trees (for trusted data access). The
immutable-nature of updates in mLSM enables us to extend it to
support lazy certification and overcome the problem of baseline
solutions (Section~\ref{sub:baseline}).

\cut{
In the rest of this section, we present the design of LSMerkle.
LSMerkle inherits some features of mLSM in its structure, and how
updates and reads are served. However, there are some key differences
in how batches of updates are represented and how LSMerkle is
extended to utilize lazy certification and to integrate in the
edge-cloud model of WedgeChain. For these reasons, we present the
full design of LSMerkle in this section without relying on describing
the base design of mLSM. We discuss mLSM more in the related work
section.
}

\begin{sloppypar}
\textbf{Structure and configuration.}
The structure of the LSMerkle tree consists of $n$
levels---structured in the same way as a LSM tree. Each level has a
threshold number of pages, \emph{i.e.}, the threshold of level $i$,
$L_i$, is $|L_i|$ pages. A page represents the updates in a block.
The first level, $L_0$, resides in memory and contains the most
recent updates to the LSMerkle tree. Each page consists of put
operations in addition to meta-information such as the range of keys
in the page and a timestamp of the time the page was created.

For the rest of this section, we assume that the number of levels is
3 and the threshold number of pages per level are: 2 for levels $L_0$
and $L_1$, and 4 for $L_2$. This is not a practical configuration,
but is chosen for ease of exposition and to simplify the presented
examples.  
\end{sloppypar}

\begin{figure}[t]
\centering
  \includegraphics[width=3.2in]{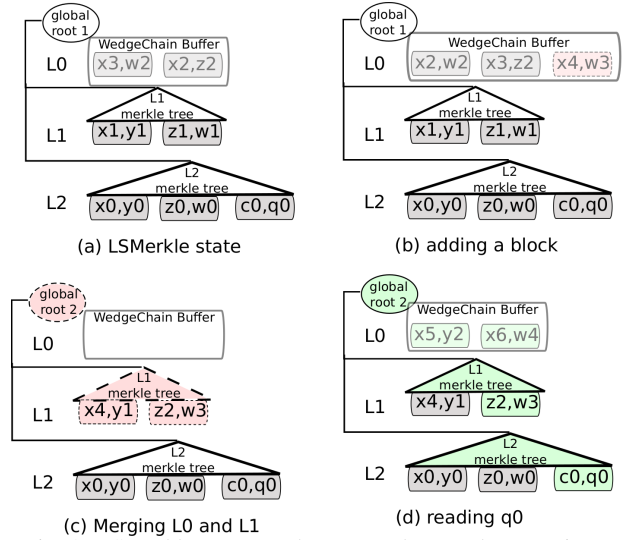}
  \caption{{\color{black}LSMerkle tree sample state and example
operations.}}
  \label{fig:lsmerkleexample}
\end{figure}

\textbf{Merklizing LSM.}
The state of a LSMerkle tree is shown in
Figure~\ref{fig:lsmerkleexample}(a). Each block in the figure
contains two put operations ($x_i$ denotes the $i^{th}$ version
of the key $x$.)
{\color{black}Level $L_0$ in LSMerkle is a WedgeChain log/buffer as described in Section~\ref{sec:wedgechain}, which allows Phase I Commit blocks of key-value operation using asynchronous trust before they are compacted by the cloud.}
The LSMerkle tree maintains a hash for each page in $L_0$. This hash
is certified by the cloud node through \textsf{block-certify} and
\textsf{block-proof} messages, similar to certifying blocks for
\textsf{add()} operations.
For every other level in the LSMerkle tree, a Merkle tree is
maintained. For example, if $L_1$ has two pages, then it has a Merkle
tree, $LSM_i$, consisting of two leaf nodes, where each leaf node
contains the hash of a page. The Merkle root of $LSM_i$ is certified
by the cloud node. LSMerkle maintains a \emph{global root} that is
the hash of all Merkle roots.

\textbf{Put operations.}
When a block is ready to be inserted in the log, it is also
added as a new page in $L_0$. The new page contains the \textsf{put}
data operations. The certification of the
block in log and the certification of the hash of its
corresponding page in $L_0$ are going to be done through the same
\textsf{block-certify} and \textsf{block-proof} message exchange with
the cloud node. Operations on pages in $L_0$ would
leverage lazy (asynchronous) certification in the same way it is
leveraged for \textsf{put} operations.

\textbf{Merging}
If the number of pages in the edge node exceeds the threshold in
$L_0$, then all pages in $L_0$ are merged with the pages in the next
level, $L_1$.  More generally, when the number of pages in level
$L_i$ exceeds $|L_i|$, the pages in $L_i$ are merged with the pages
in $L_{i+1}$.
The merge protocol is an asynchronous protocol that does not
interfere with the normal operation of the LSMerkle tree. Consider a
merge of pages from $L_i$ into pages in $L_{i+1}$.  The edge node
sends a copy of all pages {\color{black}undergoing the merge
in $L_i$ and $L_{i+1}$ and the corresponding Merkle tree hashes to
the cloud node}.
When the cloud node receives the pages, it verifies the authenticity
of the pages and their state by checking the associated certification
proofs. Then, it performs the merge of the pages, similar to how a
LSM tree merge is performed. {\color{black}The resulting merged pages replace the
old pages.}

The cloud node computes the new level's Merkle tree to derive the
corresponding Merkle root and global root. Then, the cloud node sends
back the new pages for in addition to the signed Merkle
root for the changed levels and a new signed global root.  When the edge node
receives the merge response, {\color{black}it replaces the pages undergoing the
merge with the new merged pages.} Also, the Merkle
roots and global root are updated with the received ones.

\textbf{Reading.}
Like an LSM tree, the LSMerkle might have redundant versions of the
same key. It is possible that the same key has multiple versions in
different pages in $L_0$. Also, a key might have versions in more
than one level. However, levels other than $L_0$ are guaranteed to
have at most one version of each key because the redundancies are removed in
the merge process.
The read algorithm should take these redundancies in consideration to
ensure that only the most recent version is returned. This is trivial
in regular LSM trees. However, in LSMerkle, we need to return the
most recent version in addition to a proof that it is indeed the most
recent version.

To prove that a returned version of a key is the most recent one, the
edge node must provide a proof that all pages in $L_0$ and pages in
other levels do not have a more recent version. Consider the case
when the edge node finds the most recent version in a page $p$ in $L_0$.
In this case, the edge node only needs to send the page $p$ in
addition to the other pages in $L_0$. The client checks that the
returned $p$ has the most recent version by reading the other $L_0$
pages. There is no need to return pages at other levels because even
if they contain other versions of the key, they are guaranteed to be
older versions.

Now, consider the case when the most recent version is in page $p$
in level $L_i$, for $i>0$. The edge node returns $p$. Also, it needs
to return a proof that every level $L_j$, where $j<i$, does not
contain a more recent version of the key. All pages in $L_0$ need to
be returned in the response because they all might have a more recent
version of the key than the one in $L_i$. For other levels between
$L_0$ and $L_i$, the edge node needs to return the page that has the
range that contains the key. For example, in $L_j$ ($0<j<i$), keys
are sorted across pages. Only one page has the range that include
the key. The edge node returns such page for each level between $L_0$
and $L_i$. Each page contains special values called $min$ and $max$
that denote the minimum and maximum keys in the range of that page.
We enforce that the first page has a $min$ of 0 and the last page has
a $max$ of infinity. Also, every two consecutive pages $p_x$ and
$p_y$ have the invariant that $p_x.max = p_y.min - 1$. This ensures
that a client can use the $min$ and $max$ to verify that the key it
is looking for is not in any other page in that level.

When the client receives the response from the edge node, it verifies
the authenticity of the response and that the returned version is
indeed the most recent one from the returned state. Afterwards the
read terminates. Some of the returned blocks might not have been
certified by the cloud node in the response. In this case, the read
is considered in the Phase I Commit. The client waits for the
\textsf{block-proof} to enter the Phase II Commit.

If the key does not exist, then the edge node returns the
intersecting pages from all levels in addition to all pages in $L_0$
to the client with their corresponding Merkle roots and global root.



\subsection{Example}
We now present an example of doing put, merge and read operations on
the LSMerkle in Figure~\ref{fig:lsmerkleexample}(a). The first
example is of adding a new block with new values of keys $x$ ($x_4$)
and $w$ ($w_3$). This is shown as the new page (shaded with the color
red) in Figure~\ref{fig:lsmerkleexample}(b). This new block triggers
a merge request since the number of blocks in $L0$ exceeds the
threshold of two blocks. Figure~\ref{fig:lsmerkleexample}(c) shows
the outcome of the merge operation. The edge node sends all the
blocks in $L0$ and $L1$ to the cloud node, and the cloud node
responds with the merged blocks. The edge node updates the tree by
emptying $L0$ and $L1$ and adding the received merged blocks to $L1$.
The Merkle tree for $L1$ and the global root are updated to reflect
the changes (all changes in the LSMerkle are represented with the
components shaded with the color red.) 

Figure~\ref{fig:lsmerkleexample}(d) shows an example of reading the
key $q$. (we assume that there are two blocks added in $L0$ before
the read.) Since the most recent value of the key is in $L2$, the
edge node returns the intersecting pages in both $L1$ and $L2$ along
with their corresponding partial Merkle trees. The edge node also
sends all the pages in $L0$ in addition to the signed global root.
(all the components that form the response to the read request are
shaded with the color green.)

\cut{
\subsection{LSMerkle Lazy Certification Algorithms}

\subsubsection{\textbf{Put operation Algorithms}}

\textbf{Client algorithm.}
The \textsf{put} algorithm is the same as the \textsf{add}
algorithm (Algorithm~\ref{alg:add}). The difference is that the input
$entry$ is a key-value pair annotated as being a \textsf{put}
operation.

\textbf{Edge node algorithm.}
When an edge node receives a \textsf{put} request, it processes
it in the same way as an \textsf{add} request. It buffers it and
eventually add it to the next block. And then it sends an
\textsf{add-response} to the client and a \textsf{block-certify}
request to the cloud node. Eventually, when a \textsf{block-proof} is
received from the cloud node, it is forwarded to the client. 

The difference in the case of LSMerkle and \textsf{put} requests
is that the edge node maintains the LSMerkle data structure and adds
the new block to $L_0$. 

\textbf{Cloud node algorithm.} 
The algorithm for the cloud node is the same for responding to
\textsf{block-certify} messages with \textsf{block-proof} messages.
The difference is that the cloud node also tracks whether the edge
node is sending merge requests for the $L_0$ blocks. The cloud node
detects that the edge node is not doing so if it receives a number of
pages that is larger than the threshold for $L_0$ but no merge
requests are received yet.

\subsubsection{\textbf{Merge Algorithms}}
To remove the redundant entries in the LSMerkle index, merge
algorithms compact pages when the number of pages in a level exceeds
a threshold. 

\textbf{Edge node algorithms.}
After adding a new block, the edge node checks if the threshold of
$L_0$ is exceeded. If it is exceeded, then the edge node sends all
pages in $L_0$ and $L_1$ to the cloud node as part of a
\textsf{merge-request} message. The request contains the pages and
their corresponding \textsf{block-proof} messages for pages in $L_0$
and signed Merkle roots for pages in other levels.

Then, it waits until it receives the new pages for $L_1$. In the
meantime, new blocks are still added to $L_0$. When the new $L_1$
pages are received, the edge node removes the pages that were merged
from the index and write the received pages in $L_1$. The
\textsf{merge-response} from the cloud node also includes a signed
Merkle root of the Merkle tree that represents the new pages of
$L_1$ and the new global root. The edge node constructs the Merkle tree for the new pages in
$L_1$ and maintains the corresponding signed Merkle root and global
root for future read requests. 

Once a merge of levels $L_i$ and $L_{i+1}$ is performed, the edge
node checks whether the threshold for $L_{i+1}$ is exceeded. If it
is, a new merge for levels $L_{i+1}$ and $L_{i+2}$ is performed
similar to the merge protocol above.

To simplify presentation, no concurrent merges are allowed. When a
new merge request is triggered, it first waits until the previous
merge and all cascading merge requests are done. If needed, this
restriction can be relaxed similar to how concurrent merges are
performed in LSM tree variants~\cite{luo2018lsm}.

\textbf{Cloud node algorithms.}
When a cloud node receives a \textsf{merge-request} message, it first
verifies that the requested merge is valid. Because only one merge
request is allowed at a time, the block ids of the next
merge request are predictable. A merge request is valid only if the
request matches the next prediction. For example, if the threshold of
$L_0$ and $L_1$ is 2 pages, then the first merge request must be for
blocks 0 and 1. Likewise, the second merge request must be for blocks
2 and 3, and so on. If the result of a merge exceeded the threshold
of the next level, then the cloud node knows that the next request is
going to be for the pages in that level. For example, after merging
0, 1, 2, and 3, assume that the number of new pages in $L_1$ becomes
3. In this case, the cloud node knows that the next merge request is
going to be for the three new pages in $L_1$.

After verifying that the merge request is valid, the cloud node
verifies that the pages are authentic. If the pages are in $L_0$, the
cloud node verifies their authenticity by checking their
corresponding \textsf{block-proof} messages. For other levels, the
cloud node constructs the Merkle tree and compare the derived Merkle
root with the Merkle root received from the edge node that was
previously signed by the cloud node. 

After doing all these verifications, the merge process starts. All
the entries in the sent pages are sorted and redundancies are
removed. Then, the sorted entries are divided into pages with the
same size (the page size is a configuration parameter). Each new
page is annotated with a timestamp and $min$ and
$max$ numbers that represent the boundaries and range of the page.
The first page's $min$ is always 0 and the last page's $max$ is
always infinity. Also, for any two consecutive pages $p_i$ and $p_j$,
the max of the first page, $p_i.max$ is always equal to $p_j.min-1$,
even if there are no keys on these boundaries.  This ensures that a
client can exactly know the range of a page without checking the
adjacent pages. 

Afterwards, the cloud node constructs a Merkle tree of the new pages
and signs the corresponding Merkle root. Also, it constructs a new
global root. The pages, signed Merkle root, and signed global root
are then sent back to the edge node.

\subsubsection{\textbf{Read Algorithms}}

\alglanguage{pseudocode}
\begin{algorithm2e}[t]
\caption{Client algorithm to read a key-value pair. }
\label{alg:get}
\begin{algorithmic}[1]
\State on get (in: key) \{
\State \,\,\,\,\,Send \textsf{get(\$key)} to edge node
\State \,\,\,\,\,$\$merkleValue$, $\$merkleProofSet$,
$\$merkleRootSet$, $\$globalRoot$ $\leftarrow$ edge node response
\State \,\,\,\,\,Verify that $\$merkleValue$, $\$merkleProofSet$,
$\$merkleRootSet$, and $\$globalRoot$ are authentic
\State \,\,\,\,\,\textbf{for} $\$i$: 0 \textsf{to}
$\$merkleProofSet.size$ - 1
\State \,\,\,\,\,\,\,\,\,\,Verify that $\$key$ is not in
$\$merkleProofSet[\$i]$
\State \,\,\,\,\,\textbf{if} $\$merkleValue$ is \textsf{null}
\State \,\,\,\,\,\,\,\,\,\,Verify that $\$merkleProofSet.size = |L|$
\State \,\,\,\,\,\,\,\,\,\,return \textsf{null} and exit
\State \,\,\,\,\,return value from $\$merkleValue$
\State \}
\end{algorithmic}
\end{algorithm2e}

\textbf{Client node algorithms.}
The client node algorithm to read from the LSMerkle tree is similar
to the algorithm to read a block (Algorithm~\ref{alg:read}). However,
the key difference is that the read request reads a key-value pair
and that the edge node returns a proof consisting of a set of partial
Merkle trees, a global root, and pages instead of just returning one
block. 

Algorithm~\ref{alg:get} shows the algorithm. First, a \textsf{get}
request is sent to the edge node. Then, the edge node returns the
following:
(1)~$merkleValue$: this is the partial Merkle Tree of the LSMerkle
page and level that contains the value, or \textsf{null} if the key
does not exist.
(2)~$merkleProofSet$: this a set of partial Merkle trees for levels
$L_j$ which consists of $L_0$ and levels that are lower than the
level that contains the value $L_i$ ($j < i$). If the key does not
exist, then this set contains all levels.  For levels larger than
$L_0$, not the whole Merkle tree is included.  Rather, only the page
that contains the range of the $key$ is included along with the
partial path to the corresponding Merkle root.
(3)~$merkleRootSet$: this is a set of signed Merkle roots for all
levels of the LSMerkle tree. Each one is signed by the cloud node,
which is delivered during prior certification or merge operations.
(4)~$globalRoot$: this is the global root of the LSMerkle tree that
is singed by the cloud node.

The cloud node verifies that the received response from the edge node
is authentic. It first computes the Merkle roots of all levels in
$merkleValue$ and $merkleProofSet$. Then it uses the computed Merkle
roots along with the Merkle roots of the remaining levels (from
$merkleRootSet$) to derive the global root. Finally, it compares the
derived global root with the received $globalRoot$ that is signed by
the cloud node. If they match, then the response is authentic.

The client then verifies that the most recent value of the key is in
the level of $merkleValue$. This is done by checking that the key
does not exist in lower levels by observing $merkleProofSet$. For
each level in $merkleProofSet$ the page is read to verify that the
key falls within its range and that the key does not exist in the
page. Finally, the value is returned. If the key is not in the
LSMerkle tree, then the client verifies that it checked all tree
levels.

\textbf{Edge node algorithms.}
When the edge node receives a \textsf{get} request, it reads the
LSMerkle tree to find the most recent version. If the key is not
found, then its sets $merkleValue$ to \textsf{null}. Otherwise, it
sets it to the page and partial path to the corresponding Merkle
root. The edge node also returns the proof that $merkleValue$
contains the most recent version, which consists of all pages in
$L_0$ and corresponding pages in levels between $L_0$ and the level
where the $merkleValue$ is. For each one of these levels, the page
with the range that encompasses the key is returned with the partial
Merkle tree path to the Merkle root of that level. The edge node also
sends the signed $merkleRootSet$ and $globalRoot$ as defined in the
client algorithms.
}

\subsection{Read Data Freshness}
\label{sub:freshness}

The LSMerkle algorithms ensures that the returned value is one that
has been added in the past and is part of a consistent snapshot of the
LSMerkle tree. However, LSMerkle does not guarantee that the read is
going to return the most recent value. This is because an edge node
might serve the read from a stale snapshot of the data. Enforcing
that a read would return the most recent value requires extensive
coordination between clients, edge nodes and the cloud node, which we
view as prohibitive. Alternatively, we propose a guarantee of reading
from a \emph{freshness window.} For example, a guarantee that the
read returns the state from a consistent snapshot as of a time no
longer than $X$ seconds ago.
To enforce this freshness property, the
following changes need to be applied to our algorithms:\\
(1) The cloud node timestamps the global root of each merged
LSMerkle. The signature would be of both the timestamp and the global
root.\\
(2) When a client receives a read response, it also checks the
timestamp of the received global root and verify its authenticity
using the signature. If the timestamp is within the freshness window,
the client accepts the response. Otherwise, it retries the request.

An issue may arise if updates are not happening frequently enough to
trigger updating the global root by the cloud node. In such a
case, the edge node can add \textsf{no-op} operations to trigger
merges more rapidly and reconstruct the LSMerkle tree.
%

\textbf{Effect of time synchronization bounds.}
Another issue is that of clock synchronization. Depending
on the distance between nodes, current time synchronization
technologies achieve synchronization with an accuracy of 10s to 100s
of milliseconds. This limits the use of our technique to the bounds
of time synchronization.

An alternative to the method we present above is to
maintain more state information at the client side (\emph{e.g.},
similar to client-side session consistency solutions) that
would allow a client to check whether the read state is
consistent and fresh by checking with its local client-side
information. Another alternative is to establish a secure
communication channel between the client and the cloud node to verify
freshness.




\cut{
Data is structured into keys and values. The access to the store is
via a transactional interface, where a transaction is a bundle of
read and write operations. LSMerkle guarantees that transactions are
serializable~\cite{}. A client can also issue read-only transactions,
which is a collection of read operations. The response to the
read-only transaction is going to be from a consistent snapshot.
Additionally, it will not conflict with on-going read-write
transactions. To enable this non-interference, the reads are not
guaranteed to be from the most recent snapshot---nonetheless, the
snapshot is guaranteed to be consistent. The read-only transactions
also allows time-traveling queries, where the client specifies the
\emph{as-of} time of the snapshot.   
}

\begin{table}[t]
\begin{center}
        \begin{tabular}{ c | c  c  c  c c }
        &        $C$ & $O$ & $V$ & $I$ & $M$ \\ \hline
        $C$ & 0   & 19  & 61 & 141 & 238 \\
        \end{tabular}
        \caption{The average Round-Trip Times in milliseconds between
California and other datacenters.}
        \label{tab:rtts}
\end{center}
\end{table}

\section{Experimental Evaluation}\label{sec:eval}

We present a performance evaluation of WedgeChain in this section.
To emulate edge-cloud coordination, we run our experiments across
geographically distributed Amazon AWS datacenters. In most of our
experiments, one datacenter---California ($C$)---will act as the edge
location, hosting clients and edge nodes and another
datacenter---Virginia ($V$)---will host the cloud node. We vary the
place of the edge and cloud nodes in some experiments across
datacenters in California ($C$), Virginia ($V$), Oregon ($O$),
Ireland ($I$), and Mumbai ($M$). The Round-Trip Times (RTTs) between
the four datacenters range between 19ms and 238ms
(Table~\ref{tab:rtts}.) In each datacenter we use Amazon EC2
\textsf{m5d.xlarge} machines. Each machine runs Linux and have 4
virtualized CPUs and 16 GB memory.
%

We compare with a baseline that processes all requests in the cloud
node. We call this baseline, \emph{cloud-only}. We also compare with
a baseline where all requests are certified first at the cloud and
then sent to the edge node. This is the baseline we described in
Section~\ref{sub:baseline}. We call this baseline the
\emph{edge-baseline}. The cloud-only baseline represents the case
where clients can fully trust the results since no edge nodes are
involved in processing them.  However, clients incur the wide-area
latency to the cloud for every request. The edge-baseline, on the
other hand, represents the current way of utilizing untrusted nodes
for data access, where data is certified first in the trusted node
(the cloud node in our case) and then sent to the untrusted node (the
edge node in our case.)
{\color{black}In the edge-baseline implementation we tested with both a vanilla Merkle Tree as well as a mLSM as the trusted index component. The choice of the index did not have a significant effect on performance as the edge-cloud coordination dominated the performance overhead. The results shown are for using mLSM as the index in edge-baseline.}

\cut{
WedgeChain handles two
types of workloads: (1)~Edge data: this is data that represents
arbitrary information that could be generated from IoT and edge
devices. This data is added to the log but does not affect the
indexing structure. (2)~Data operations: this is data in the form of
\textsf{put} and \textsf{get} operations.  This workload affects the
indexing components of WedgeChain.
}
We use key-value operations 
in our experiments since it affects both logging and indexing
components. We batch \textsf{add} and \textsf{put} requests in all
experiments. Unless mentioned otherwise, each batch consists of 100
\textsf{put} operations, and the size of the value of each operation
is 100 bytes.  Each edge node maintains one partition of the data,
which consists of 100,000 key-value pairs. In the experiments, we
report the performance of one partition.
The LSMerkle tree has four levels. The thresholds for the levels are
10, 10, 100, and 1000 pages for levels 0, 1, 2, and 3, respectively.

\begin{figure}[!t]
\begin{center}
\subfigure[Latency]
{\label{fig:batch_latency}
\includegraphics[scale=.63]{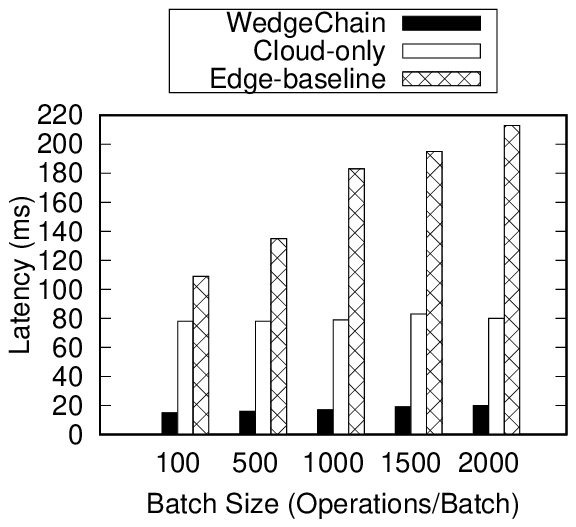}}
\subfigure[Throughput]
{\label{fig:batch_throughput}
\includegraphics[scale=.63]{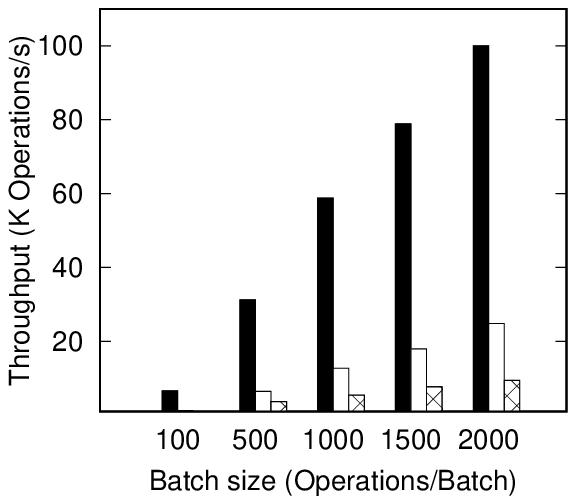}}
\caption{The performance of Put operations while varying the block size}
\label{fig:batching}
\end{center}
\end{figure}

\subsection{Baseline Performance of Put Operations}

\cut{
We start the evaluation study with a baseline experiment that
compares WedgeChain with the two baselines: cloud-only and
edge-baseline. Figure~\ref{fig:batching} shows the results of this
experiment.
}
In Figure~\ref{fig:batching}, we vary the batch (\emph{i.e.}, block)
size from 100 to 2000 operations per batch.
Figure~\ref{fig:batch_latency} shows the latency results. WedgeChain
achieves the lowest latency, which is below 20ms. This latency
corresponds to the Phase I Commit latency. This low latency is
expected since it is the time needed to communicate with the nearby
edge node. On the other hand, Cloud-only incurs a latency around 80ms
which corresponds roughly to the round-trip time from California to
Virginia in addition to the processing overhead. Edge-baseline also
incurs a cost due to having to wait for the response from the cloud
which leads to a latency higher than 100ms in all cases.
Edge-baseline performs worse than Cloud-only due to the need to
involve the edge node in the commitment of the block which requires
more time. 

As we increase the batch size, the latency increases slightly for
WedgeChain from 15ms to 20ms and Cloud-only from 78ms to 83ms.
However, Edge-baseline is affected significantly by the increase in
the batch size resulting in increasing the latency from 109ms to
213ms. The reason for this increase is that both the edge node and
cloud node are involved in the commitment of the block and are in the
path of execution, which leads to stressing the network bandwidth
resources faster than the other two approaches. WedgeChain masks the
effect of this edge-cloud coordination by utilizing the concept of
lazy (asynchronous) certification that removes the cloud node from
the path of execution when adding blocks or performing \textsf{put}
operations.

Throughput results are shown in Figure~\ref{fig:batch_throughput}.
Due to WedgeChain's low latency, throughput increases
from 6.6K operations/s to roughly 100K operations/s, which is a 15x
increase that results from multiplying the batch size 20 times (from
100 to 2000 operations per block.) Cloud-only's throughput
experiences a 18.5x increase when varying the batch size from 100 to
2000 operations. The poor latency performance of Edge-baseline causes
the throughput to scale poorly, where the throughput only doubles
when increasing the block size from 100 to 2000 operations.

\cut{
\begin{table}[t]
\begin{center}
        \begin{tabular}{ | l | c | }
	\hline
        &        Latency (ms) \\ \hline
        WedgeChain & $<$1ms \\ \hline
	Cloud-only & 62ms \\ \hline
	Edge-baseline & $<$1ms \\ \hline
        \end{tabular}
        \caption{Read latency with a client and edge node in
California and a cloud node in Virginia.}
        \label{tab:read}
\end{center}
\end{table}

Table~\ref{tab:read} shows the performance of read operations. Both
WedgeChain and Edge-baseline utilize edge nodes close to the users to
serve reads. This makes reads fast---incurring only the latency between the
client and the edge node, which is in this case below 1ms.
Cloud-only, on the other hand, forces read requests to be served from
the cloud node and thus incur the round-trip time latency to the
cloud node, which leads to a read latency of 62ms.
}

{\color{black}
\begin{figure}[!t]
\begin{center}
\subfigure[\color{black}All-write workload]
{\label{fig:multiclient_allwrite}
\includegraphics[scale=.63]{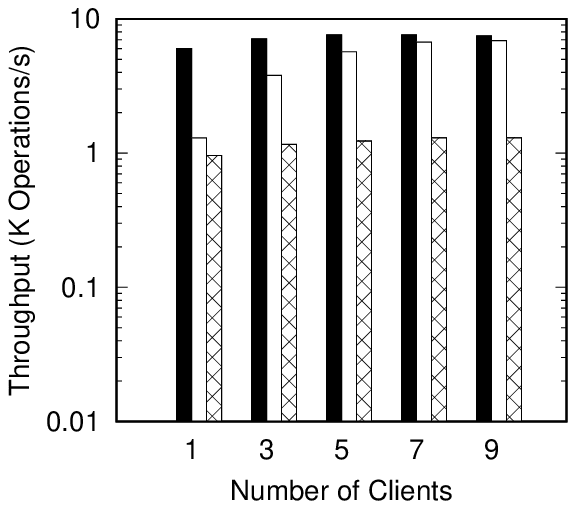}}
\subfigure[\color{black}50\% reads, 50\% writes]
{\label{fig:multiclient_half}
\includegraphics[scale=.63]{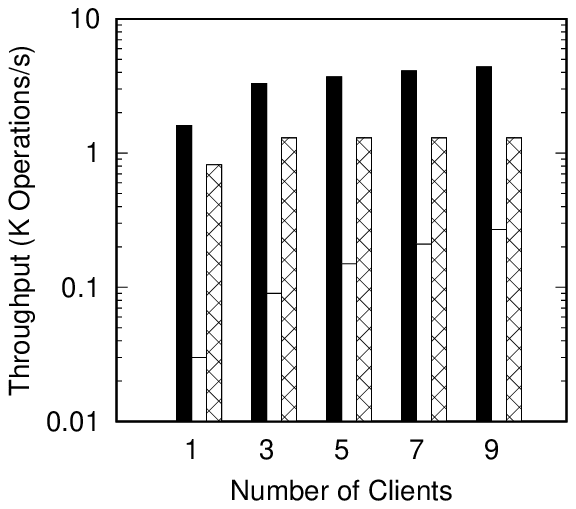}}
\subfigure[\color{black}All-read workload]
{\label{fig:multiclient_allread}
\includegraphics[scale=.63]{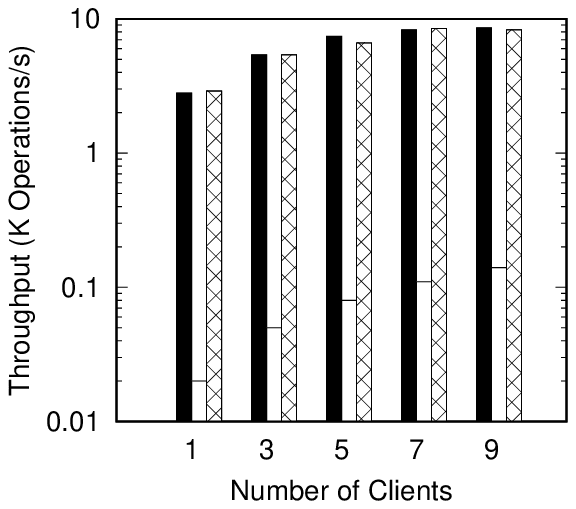}}
\subfigure[\color{black}Best-case read latency and verification
overhead]
{\label{fig:multiclient_readbatch}
\includegraphics[height=1.2in,width=1.6in]{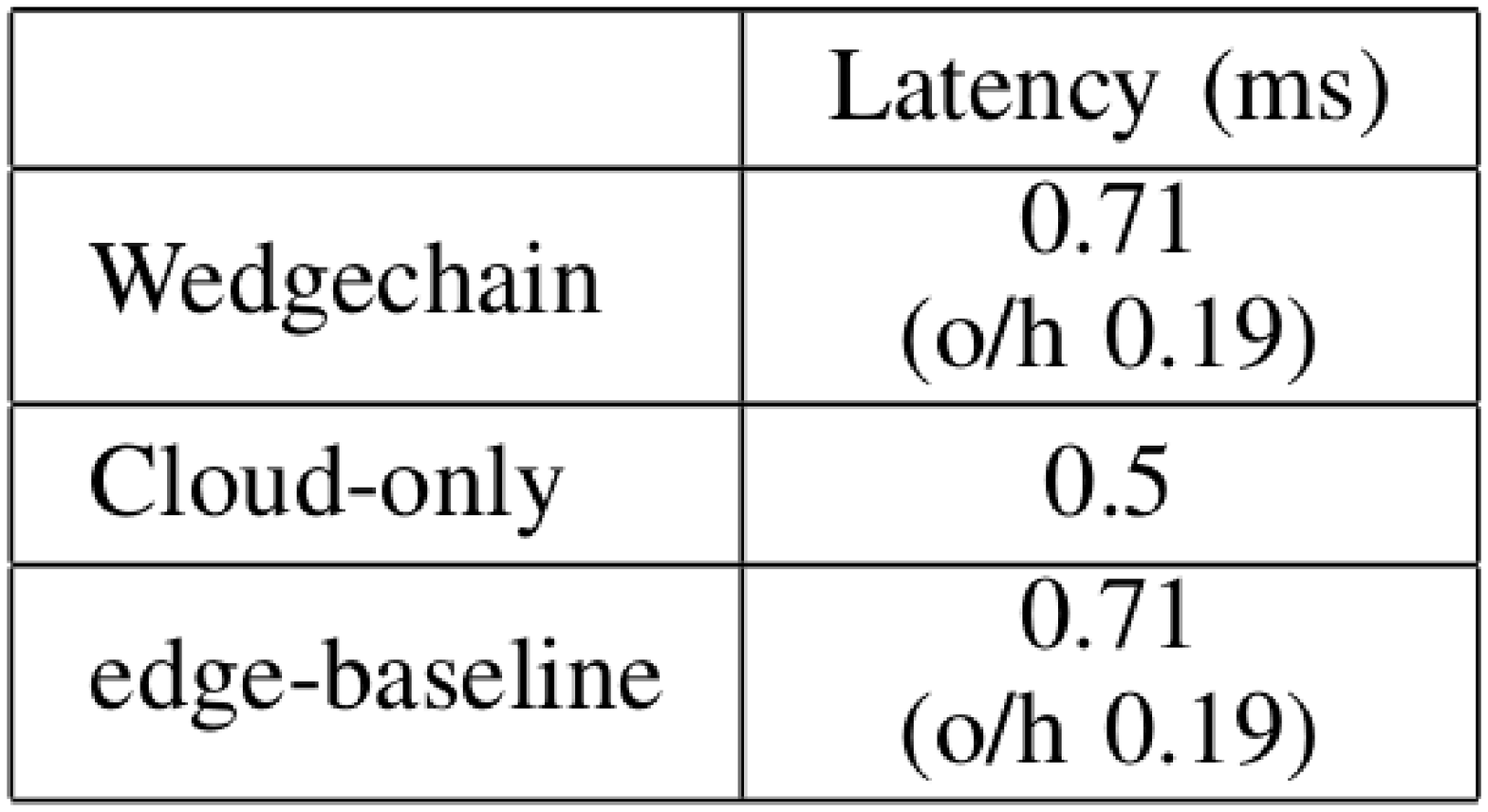}}
%
\caption{\color{black}Results of multi-client and mixed workload experiments}
\label{fig:multiclient}
\end{center}
\end{figure}

\subsection{Multi-Client and Mixed Workload}
\label{sub:multiclient}

Figure~\ref{fig:multiclient} shows experiment results while varying
the number of clients and read-to-write ratio.
Figure~\ref{fig:multiclient_allwrite} presents experiments with an
100\%-write workload while varying the number of clients from 1 to 9.
Increasing the number of clients allows more concurrency. This leads
to an increase in throughput by 22--30\% for WedgeChain and
Edge-baseline. However, the increase for Cloud-only is much higher at
433\% which enables Cloud-only to catch up to WedgeChain throughput
and be only 7\% lower than the throughput of WedgeChain. The main
reason for this is that by increasing the number of clients,
Cloud-only is offsetting the overhead of communication (edge-cloud
latency) while not incurring the communication and computations
overheads of WedgeChain. 
On the other hand, Edge-baseline suffers from the synchronous
communication overhead which causes its latency to be the lowest out
of the three.

Figure~\ref{fig:multiclient_half} shows experiments with a mixed
workload of 50\% reads and 50\% writes. In this experiment, writes
are buffered, but reads are interactive (we show a case that stresses
read performance later in Figure~\ref{fig:multiclient_readbatch}).
Interactive reads are very expensive for Cloud-only as each read
requires the client to wait for the response that takes a duration
proportional to the edge-cloud latency. This causes the throughput of
Cloud-only to reach only 270 operations/s while WedgeChain achieves
4K operations/s and Edge-baseline achieves 1.3K operations/s. The
reason for Edge-baseline achieving a lower performance than
WedgeChain is due to the 50\% writes that incur the synchronous
coordination overhead.

Figure~\ref{fig:multiclient_allread} shows experiments with a
read-only workload. In this experiment, WedgeChain and Edge-baseline
perform similarly. In both solutions, interactive read operations
involve the same steps of communication and verification for
WedgeChain and Edge-baseline. However, Cloud-only requires
communication with the cloud and since interactive reads requires the
client to wait until the read is served, this leads to a high
overhead and achieving a small fraction of the performance of
WedgeChain and Edge-baseline. 

The significant difference between the read performance of WedgeChain
and Edge-baseline on the one hand and Cloud-only is due to the
communication latency. However, Cloud-only does not incur the complex
computations needed in edge nodes since its results are trusted. This
would enable the cloud node to process more reads than what the edge
nodes can process. Observing this while increasing the number of
clients is challenging as it would require emulating a huge number of
clients. Instead, we perform an experiment in
Figure~\ref{fig:multiclient_readbatch} where we measure the best-case
read latency directly at the cloud node for Cloud-only and at the
edge node for the others. WedgeChain and Edge-baseline achieves a
similar read performance of 0.71ms, 0.19ms of which is due to the
verification overhead performed at the client. Cloud-only, on the
other hand, achieves a better latency of 0.5ms without incurring a
verification overhead as the results are trusted. 

}

\subsection{Asynchronous Certification and Commit Phases}

\begin{figure}[!t]
\begin{center}
\includegraphics[scale=.63]{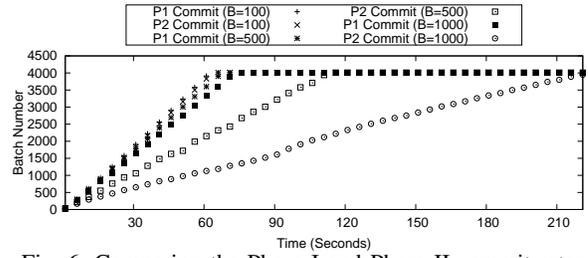}
\caption{Comparing the Phase I and Phase II commit rates.}
\label{fig:scaling}
\end{center}
\end{figure}

The performance advantage of WedgeChain is due in a large part to the
concept of lazy (asynchronous) certification that distinguishes
between two phases of commitment. In this section, we provide more
insights about lazy certification by showing the relation between the
two phases of commitment. Figure~\ref{fig:scaling} shows the results
of three experiments, each for a different batch size. In each
experiment, WedgeChain commits 4000 batches (blocks). The figure
shows how rapidly the batches are being committed by plotting the
number of committed batches against the x-axis that represents time.  

For the case of 100 operations per batch (B=100), the rates of both
Phase I Commit (P1) and Phase II Commit (P2) are similar---the two
plots are overlapping and the 4000 batches are committed within 60
seconds.  This means that although Phase II Commit takes more time,
it is happening at the same rate of the Phase I Commit.
This is different when we start increasing the batch size. For the
case of 500 operations/batch (B=500), there is a difference in the
rate of P1 and P2 commits. P1 commit is still fast; committing the
4000 batches within 60s. However, P2 commits take more than 100s. The
reason for this is that the buffering and processing of larger batch
sizes lead the P2 throughput to be lower than the P1 throughput.
This is the same case with larger batch sizes, such as the third
experiment with 1000 operations/batch (B=1000).

The main takeaway of this set of experiments is the behavior of P1
and P2 commits and how WedgeChain is able to mask both the latency
and throughput overhead of edge-cloud communication. Notice that in
all cases, the P1 commit is still able to commit the 4000 batches in
close to 60s, even if P2 commit takes much longer. This is the
feature of lazy certification that we desire, which is masking the
overhead of expensive edge-cloud coordination and enjoying the low
latency and high performance that can be delivered by the nearby edge
node.

\subsection{The Effect of Edge-to-Cloud and Client-to-Edge Latency}

\begin{figure}[!t]
\begin{center}
\subfigure[Varying the cloud node location]
{\label{fig:rtt_cloud_latency}
\includegraphics[scale=.63]{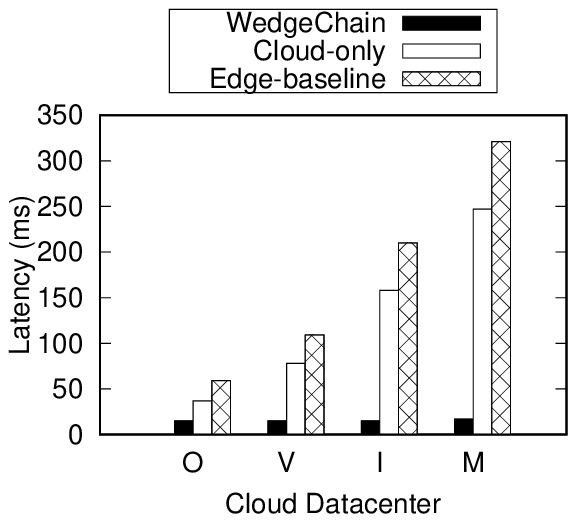}}
\subfigure[Varying the edge node location]
{\label{fig:rtt_edge_latency}
\includegraphics[scale=.63]{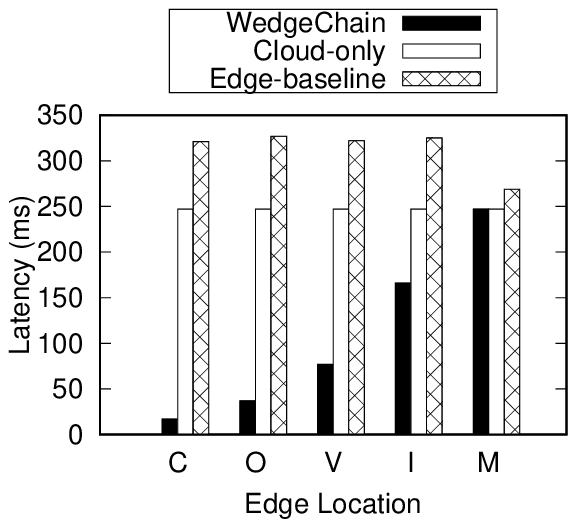}}
\caption{The latency of committing blocks of Put operations while varying the location
of the cloud and edge nodes.}
\label{fig:rtt_cloud}
\end{center}
\end{figure}

The performance advantage of WedgeChain relies on the close proximity
of clients and edge nodes, which masks the wide-area latency between
the clients and the cloud node. Therefore, the magnitude of the
performance advantage depends on the relative communication latency
between the client and the edge node on one hand, and the
communication latency between the edge node and the cloud node on the
other hand. To measure the effects of these relative latencies, we
performed two sets of experiments that vary the locations of the edge
and cloud nodes.

Figure~\ref{fig:rtt_cloud_latency} shows the latency while varying
the location of the cloud node and fixing the locations of the client
and edge node in California. WedgeChain is able to preserve the
latency benefit of utilizing an edge node---the latency in all cases
is within 15ms and 17ms. This shows that it successfully masks the
wide-area latency even in cases when the round-trip time to the cloud
node is 238ms (between California and Mumbai). Cloud-only is affected
by the location of the cloud node since all requests are served by
the cloud node. The latency for Cloud-only ranges between 37ms and
247ms, which corresponds to the round-trip latency to the cloud node.
This is also the case for Edge-baseline where the latency ranges
between 59ms and 321ms. Edge-baseline performs worse than Cloud-only
due to the bandwidth and computation stress incurred to synchronously
coordinate between the edge and cloud nodes as we observed in the
previous experiments. 

Figure~\ref{fig:rtt_edge_latency} varies the location of the edge
node while fixing the client in California and the cloud node in
Mumbai. WedgeChain latency is affected directly by the location of
the edge node, since that is where requests are committed.
WedgeChain's latency vary between 17ms (when the edge node is in
Oregon) and 247ms (when the edge node is in Mumbai). The latency
in all cases corresponds to the round-trip latency from the client to
the edge node.
Cloud-only does not utilize an edge node and thus experiences the same
performance in all cases, which corresponds to the latency from the
client to the cloud node.
Edge-baseline incurs a similar latency in all cases while varying the
location of the edge node except for the case when the edge node is
co-located with the cloud node. The reason for the similarity in all
cases except an edge in Mumbai is because the sum of the latencies
between the client, the edge, and cloud nodes are similar. This makes
the total time spent for communication be similar. Additionally, in
these cases, an additional overhead is incurred for edge-cloud
coordination as we observed in the evaluations above. The reason for
achieving a better performance when the edge node is co-located with
the cloud node is that the overhead of edge-cloud coordination
diminishes, leaving the communication cost to be the only dominating
cost for latency. This is why the latency of Edge-baseline is similar
to both Cloud-only and WedgeChain.

In all cases, WedgeChain outperform both Cloud-only and edge-baseline
except for the case when the edge node is co-located with the cloud
node. When the edge node is co-located with the cloud node, all three
systems perform similarly. 

{\color{black}
\subsection{Dataset Size}
\label{sub:dataset}

Here, we vary the size of the key range from
100K to 100M keys. Although we target edge-cloud environments where
we expect that edge partitions would be small, we perform this
evaluation to test the effect of the size of the partition.
Increasing to 100M keys, we do not observe a significant effect on
write performance. WedgeChain achieves a latency between 15--16ms,
Edge-baseline achieves a latency between 88--95ms, and Cloud-only
achieves a latency of 78--79ms across all cases. 
The reason for this is that the communication and verification
overheads (in the order
of 10s of milliseconds) outweigh the potential
I/O overhead caused by increasing the database size (in the order of
milliseconds or less).

}



\cut{
\vspace{-0.1in}
\subsection{LSMerkle Performance}


\begin{figure}[!t]
\begin{center}
\includegraphics[scale=.63]{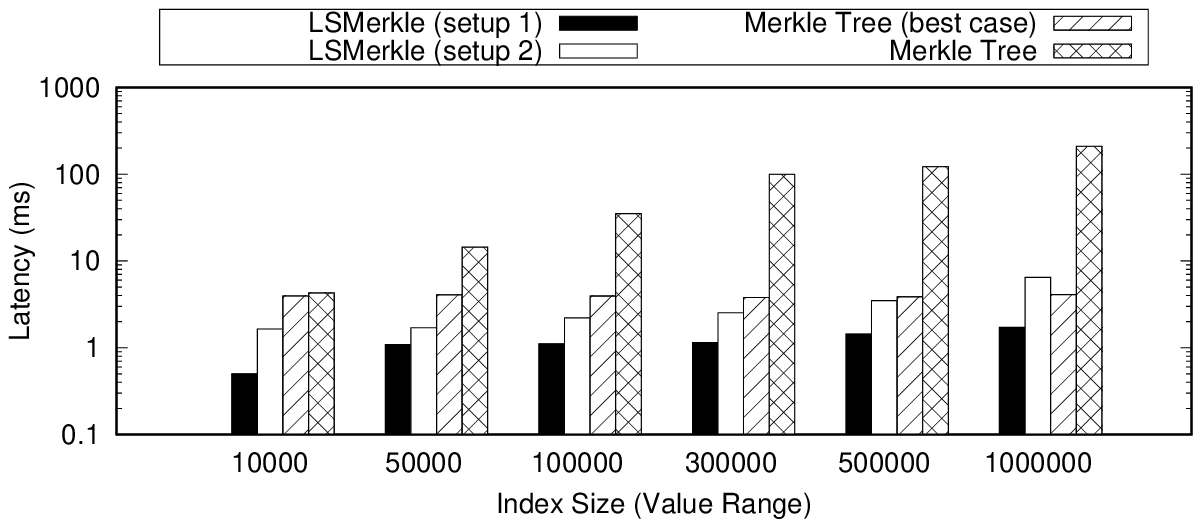}
\caption{The performance (latency to commit a block of 100
operations) of LSMerkle indexing compared with a Merkle Tree.}
\label{fig:lsmerkle}
\end{center}
\vspace{-0.2in}
\end{figure}

In this section, we present the results of an experimental evaluation
of LSMerkle. Figure~\ref{fig:lsmerkle} shows the results of this
evaluation. In this set of experiments, we test with two setups of
LSMerkle. The first (setup 1) sets the thresholds to 50 blocks for
Level 0 (L0) and L1, and 500 blocks for L2. Setup 2 sets the
thresholds to 10 blocks for L0 and L1, and 100 blocks for L2. In both
setups, L3 is the last level and has no bound. LSMerkle's performance
is compared with the Merkle Tree. We compare with two setups for the
Merkle tree. The regular one updates the Merkle tree once for every
block, where the relevant paths in the Merkle tree is recomputed---to
account for new and deleted items throughout all the changed pages.
The second setup is a \emph{best case} scenario, where we assume that
no additions and deletions are happening in the Merkle tree,
so that the only operation that is required for each update is just
the computation of the hashes from the modified page to the root. 

In this set of experiments, we attempt to isolate the performance of
LSMerkle indexing from the rest of the mechanics of WedgeChain.
Therefore, we perform the experiments on LSMerkle and the Merkle tree
by adding blocks of data directly to the data structures. The
experiment measures the latency of committing a block of 100
operations to the data structure while varying the range of the data.
We collect the latency for 3000 committed blocks while the index is
at 50\% capacity, \emph{i.e.}, there are key-value pairs for half the
key range.  

The regular Merkle tree is performing the worst due to the high
overhead of reconstructing the Merkle tree continuously. This cost
increases for larger index sizes because the number of hashing
operations becomes higher. Specifically, the latency increases from
4ms when the index size is 10000 keys to 210ms when the index size is
1,000,000 keys. Merkle tree (best case) achieves a much better latency
around 4ms for all cases. This performance represents the best-case
scenario which involves updating the paths to the root with no
restructuring of the tree due to add or remove operations. LSMerkle 
setup 1 (LSMerkle 1), achieves a latency between 0.5ms and 1.72ms for
all cases. The Merkle tree best case latency is 2.4--8x higher than
LSMerkle 1. LSMerkle 2 achieves a latency between 1.65ms and 6.48ms
in all cases. LSMerkle 2 is slower that LSMerkle 1 because the setup
with smaller thresholds causes merge operations to be more frequent
and thus incur more overhead. 

To summarize, WedgeChain successfully masks the wide-area edge-cloud
latency overhead and achieves a performance comparable to running
purely on edge resources close to users. Also, LSMerkle outperforms
Merkle trees by enabling a fast-ingestion of data and consolidating
data asynchronously.

\vspace{-0.1in}
}


\section{Related Work}\label{sec:related}

\begin{sloppypar}
\cut{
WedgeChain is at the intersection of three areas: wide-area data
management, edge-cloud systems, and trusted coordination (byzantine
agreement).

Wide-area data management is the study of data management systems
that span large geographic
locations~\cite{spanner,mdcccc,paxoscp,replicatedcommit,nawab2015minimizing,PNUTS,COPS,bailis2013bolt,lloyd2013stronger,chariots,walter}.
Like WedgeChain, these solutions build data access systems over a
logging (SMR) infrastructure and tackle the performance challenges
due to wide-area coordination. WedgeChain is different from general
wide-area data management in that it tackles the performance
asymmetry and trade-offs exhibited in edge-cloud systems.
}

Edge-cloud data management is the area of utilizing edge
nodes~\cite{satyanarayanan2009case,bonomi2012fog} that
are closer to user to perform data management tasks to augment
existing cloud
deployments~\cite{dpaxos,nawab2018nomadic,lin2007enhancing,saxena2015edgex,eyal2015cache}
This area is also related to early work in mobile data
management~\cite{perez2007consistent,gray1996dangers,terry2008replicated}. 
WedgeChain shares the goal of utilizing edge resources for data
management. The distinguishing feature of WedgeChain compared to
this set of work is that it considers a byzantine fault-tolerance
model where the edge nodes are not trusted.

Coordination with untrusted nodes (Byzantine
fault-tolerance)~\cite{lamport1982byzantine,pease1980reaching} has
been investigated extensively in the context of data
systems~\cite{castro1999practical,kotla2007zyzzyva,cowling2006hq,
abd2005fault,kotla2004high,yin2003separating,li2007beyond,amir2010steward}
and
databases~\cite{DBLP:journals/sigmod/BargaL02,luiz2014mitra,neiheiser2018fireplug},
and is recently gaining renewed interest due to emerging blockchain
and cryptocurrency
applications~\cite{androulaki2018hyperledger,quorum,chain,parity,ripple,DBLP:conf/ndss/Al-BassamSBHD18}.
WedgeChain contribution to this body of work is
(1)~the introduction of
the concept of lazy (asynchronous) certification to byzantine
fault-tolerance. Existing byzantine fault-tolerance protocols require
extensive communication and coordination to prevent malicious
activity, which makes them infeasible in real scenarios. Lazy
certification makes a shift from a paradigm of ``preventing'' malicious activity
that is expensive, to a paradigm of ``detect and eventually punish''
that allows better performance.
(2)~our work also tackles the unique challenges that
arise from edge-cloud systems such as a hybrid trust model,
edge-cloud trusted indexing, and asymmetric coordination.

WedgeChain's index, LSMerkle, builds on mLSM~\cite{raju2018mlsm} that
combines features of both LSM trees~\cite{o1996log,luo2018lsm} and
Merkle trees~\cite{DBLP:conf/sp/Merkle80} to produce a fast-ingestion
trusted data index. The choice of building on mLSM is due to its
append-only and immutable nature that makes it amenable to be
integrated into WedgeChain's lazy certification method.
{\color{black}LSMerkle uses mLSM as the data structure at the edge and builds an asynchronous (lazy) certification protocol around it to enable coordinating updates and compaction with the trusted cloud node. LSMerkle also integrates a WedgeChain log/buffer as a replacement to the memory component of mLSM to enable incoming key-value requests to be Phase I Committed with lazy (asynchronous) certification.}
If mLSM
is used with no changes in an edge-cloud environment, then it would
resemble the baseline (Sections~\ref{sub:baseline}) in that each
\textsf{put} operation must go to the cloud node first before being
part of the state in the edge nodes. LSMerkle, on the other hand,
allows lazy certification where \textsf{put} operations can be Phase
I Committed on the (untrusted) edge nodes without involving the
(trusted) cloud node. The same is true for \textsf{get} operations
where LSMerkle modifies the protocol for \textsf{get} operations to
allow reading Phase I committed data. To allow these extensions, {\color{black} LSMerkle builds a protocol around mLSM to perform \textsf{put}, \textsf{get}, and
\textsf{merge} operations. This protocol enables a pattern of Phase I committing locally at the edge and then
coordinating with the cloud for Phase II commit.}

\cut{
WedgeChain's index, the LSMerkle, combines features of both LSM
trees~\cite{o1996log,luo2018lsm} and Merkle trees~\cite{DBLP:conf/sp/Merkle80} to produce a fast-ingestion
trusted data index. This presents a step forward to Authenticated
Data
Structures (ADS)~\cite{DBLP:conf/sp/Merkle80,DBLP:conf/esa/Tamassia03,DBLP:conf/icde/JainP13,DBLP:conf/sp/ZhangGKPP17,DBLP:conf/ccs/ZhangKP15,DBLP:conf/popl/MillerHKS14}
in general as a new ADS that focuses on fast ingestion of data.
}

\end{sloppypar}

\vspace{-0.05in}
\section{Conclusion}
\label{sec:conclusion}
\vspace{-0.05in}
\begin{sloppypar}
WedgeChain is an edge-cloud system that tolerates malicious edge nodes.
WedgeChain's main innovations are (1)~a lazy (asynchronous)
certification strategy. Lazy certification allows edge
nodes to lie---however, it also guarantees that a lie is going to be
discovered. With proper penalties when malicious acts are discovered,
the guarantee of eventually catching the lie would deter edge
nodes from acting maliciously. (2)~WedgeChain takes the
trusted cloud node out of the execution path and minimizes edge-cloud
coordination using data-free coordination. (3)~We propose the
LSMerkle tree that extends mLSM~\cite{raju2018mlsm} to support
trusted fast-ingestion indexing while utilizing WedgeChain's features
of lazy certification and data-free coordination.
%
\end{sloppypar}

\section{Acknowledgments}
This research is supported in part by the NSF under grant CNS-1815212.



\bibliographystyle{abbrv}
\bibliography{citations,citations2,diss}


\end{document}